\pdfoutput=1
\RequirePackage{ifpdf}
\ifpdf 
\documentclass[pdftex]{sigma}
\else
\documentclass{sigma}
\fi

\def\ini{\textrm{in}}
\def\fin{\textrm{f\/in}}

\def\f{\frac}
\def\lp{{\ell}_{\rm Pl}}

\def\dd{\textrm{d}}
\def\d{\textrm{d}}

\newcommand{\C}{\mathbb{C}}
\def\tr{{\rm Tr \,}}
\def\R{\mathbb{R}}
\def\H{\mathcal{H}}

\newcommand{\bra}[1]{\ensuremath{\langle#1|}}
\newcommand{\ket}[1]{\ensuremath{|#1\rangle}}

\newcommand{\bek}[3]{{\langle#1\,|\,#2\,|\,#3\rangle}}

\newcommand{\pp}[2]{\frac{\partial^2 #1}{\partial {#2}^2}}

\newcommand{\Feyn}[1]{#1\kern-0.65em/}
\def\bc{\begin{center}}
\def\ec{\end{center}}

\def\vol{{\mathtt{v}}}

\newcommand{\fracs}[2]{{\scriptstyle\frac{#1}{#2}}} 

\newcommand{\N}{{\mathbb N}}

\newcommand{\pP}{{\mathbb P}}

\newcommand{\cE}{{\mathcal E}}

\newcommand{\cL}{{\mathcal L}}
\newcommand{\cH}{{\mathcal H}}
\newcommand{\dcH}{{{}^2{\mathcal H}}}
\newcommand{\cHi}{{{}^2{\mathcal H}_{\rm inv}}}

\newcommand{\cM}{{\mathcal M}}

\newcommand{\cC}{{\mathcal C}}

\newcommand{\SU}{\mathrm{SU}}
\newcommand{\SL}{\mathrm{SL}}

\newcommand{\U}{\mathrm{U}}

\newcommand{\vJ}{\vec{J}}
\newcommand{\vV}{\vec{V}}

\newcommand{\id}{\mathbb{I}}

\newcommand{\su}{{\mathfrak su}}
\renewcommand{\sl}{{\mathfrak sl}}

\renewcommand{\u}{{\mathfrak u}}
\newcommand{\la}{\langle}
\newcommand{\ra}{\rangle}

\newcommand{\tl}{\widetilde}

\def\eps{\epsilon}

\newcommand{\hh}{{\mathbf h}}

\newcommand{\bz}{\overline{z}}

\def\pp{\partial}

\def\te{\tl{e}}

\def\Ea{E^{(\alpha)}}
\def\Eb{E^{(\beta)}}
\def\Ua{U^{\alpha}}
\def\Ub{U^{\beta}}

\def\Fa{F^{(\alpha)}}
\def\Fb{F^{(\beta)}}

\def\bz{\bar{z}}
\def\bQ{\bar{Q}}
\def\inv{{\textrm{Inv}}}
\def\cHNJ{\cH_N^{(J)}}
\def\cHN{\cH_N}
\def\tU{{{}^t U}}
\def\vcC{\vec{{\cal C}}}
\def\cHQJ{\cH^{(Q)}_J}

\def\wM{\widehat{M}}
\def\wQ{\widehat{Q}}
\def\wbQ{\widehat{\bar{Q}}}
\def\wcC{\widehat{\cC}}

\numberwithin{equation}{section}

\begin{document}

\allowdisplaybreaks

\renewcommand{\thefootnote}{$\star$}

\renewcommand{\PaperNumber}{015}

\FirstPageHeading

\ShortArticleName{Learning about Quantum Gravity with a Couple of Nodes}

\ArticleName{Learning about Quantum Gravity\\ with a Couple of Nodes\footnote{This
paper is a contribution to the Special Issue ``Loop Quantum Gravity and Cosmology''. The full collection is available at \href{http://www.emis.de/journals/SIGMA/LQGC.html}{http://www.emis.de/journals/SIGMA/LQGC.html}}}

\Author{Enrique F.~BORJA~$^{\dag^1 \dag^2}$, I\~naki GARAY~$^{\dag^1 \dag^3}$ and Francesca VIDOTTO~$^{\dag^4 \dag^5}$}

\AuthorNameForHeading{E.F.~Borja, I.~Garay and F.~Vidotto}

\Address{$^{\dag^1}$~Institute for Theoretical Physics III, University
of Erlangen-N\"{u}rnberg, \\
\hphantom{$^{\dag^1}$}~Staudtstra{\ss}e 7, 91058 Erlangen, Germany}
\EmailDD{\href{mailto:efborja@theorie3.physik.uni-erlangen.de}{efborja@theorie3.physik.uni-erlangen.de}, \href{mailto:igael@theorie3.physik.uni-erlangen.de}{igael@theorie3.physik.uni-erlangen.de}}

\Address{$^{\dag^2}$~Departamento de F\'{\i}sica Te\'{o}rica and IFIC, Centro Mixto
Universidad de Valencia-CSIC, \\
\hphantom{$^{\dag^2}$}~Facultad de F\'{\i}sica, Universidad de
Valencia, Burjassot-46100, Valencia, Spain}

\Address{$^{\dag^3}$~Departamento de F\'{\i}sica Te\'{o}rica, Universidad del Pa\'{\i}s Vasco,\\
\hphantom{$^{\dag^3}$}~Apartado 644, 48080 Bilbao, Spain}

\Address{$^{\dag^4}$~Laboratoire de Physique Subatomique et de Cosmologie,\\
\hphantom{$^{\dag^4}$}~53 rue des Martyrs,  38026 Grenoble, France}

\Address{$^{\dag^5}$~Centre de Physique Th\'eorique de Luminy,
     Case 907, 13288 Marseille, France}
\EmailDD{\href{mailto:vidotto@cpt.univ-mrs.fr}{vidotto@cpt.univ-mrs.fr}}

\ArticleDates{Received October 08, 2011, in f\/inal form March 12, 2012; Published online March 25, 2012}

\Abstract{Loop Quantum Gravity provides a natural truncation of the inf\/inite degrees of freedom of gravity, obtained by studying the theory on a given f\/inite graph. We review this procedure and we present the construction of the canonical theory on a simple graph, formed by only two nodes. We review the $\U(N)$ framework, which provides a powerful tool for the canonical study of this model, and a  formulation of the system based on spinors. We consider also the covariant theory, which permits to derive the model from a more complex formulation, paying special attention to the cosmological interpretation of the theory.}


\Classification{83C27; 83C45; 83C60; 83F05}

\renewcommand{\thefootnote}{\arabic{footnote}}
\setcounter{footnote}{0}

\section{Introduction to \emph{few-nodes} models}

It's a long way to quantum gravity, and the way is not unique. We do not yet have a complete quantization of the gravitational f\/ield. Dif\/ferent proposals have been explored, and we are assisting to the convergence of some of them into a unique coherent picture, that takes the name of Loop Quantum Gravity (LQG) \cite{Rovelli,Rovelli:2011eq,Thiemann}.

The state space of LQG, ${H}_{\rm LQG}$, admits subspaces that are determined by graphs  $\Gamma$, whose physical meaning we discuss below.   In \cite{Rovelli:2008ys} the idea was put forward to study the truncation of the full quantum theory on a very simple graph: a graph formed by only two nodes. This truncation, it was argued, can be suf\/f\/icient to study cosmology. The idea has since been developed in various directions. First, the physical approximation involved in this truncation has become more clear. Second, the relation between the degrees of freedom captured by this ``dipole'' graph and the degree of freedom of Bianchi~XI has been clarif\/ied.

More importantly, the ``dipole'' truncation has proven to be a natural context for developing the $\U(N)$ formalism, a powerful mathematical language for controlling the mathematical structure of the quantum states of geometry, especially in the homogeneous and isotropic context, and to suggest the form of the Hamiltonian.

Finally, the ``dipole'' graph has represented the starting point for deriving cosmological amplitudes from the covariant spinfoam theory, opening the way to the use of richer graphs.

In this article we review  these dif\/ferent directions of research opened by the study of the ``dipole'' graph.  We begin, below, by discussing the physical meaning of the graph.  We discuss the original Hamiltonian quantization of dipole in the context of cosmology in Section~\ref{section2}, then the $\U(N)$ formalism in Section~\ref{section3}, and f\/inally the spinfoam application in Section~\ref{section4}.

\subsection{Why graphs?}

Let us begin by discussing how a truncation can appear in quantum gravity, and how it is related to graphs.

{\bf Discrete gravity (1961).} The essential idea behind the graph truncation can be traced to Regge calculus \cite{Regge:1961px}, which is based on the idea of approximating spacetime with a triangulation, where the metric is everywhere f\/lat except on the triangles.  On a f\/ixed spacelike surface, Regge calculus induces a discrete 3-geometry def\/ined on a 3d triangulation, where the metric is everywhere f\/lat except on the bones. The two-skeleton of the dual of this 3d cellular decomposition is a graph $\Gamma$, obtained by taking a point (a ``node'') inside each cell, and connecting it to the node in an adjacent cell by a link,  puncturing the triangle shared by the two cells.  These are the graph we are considering here.  More precisely, we will consider some generalizations of this construction, where the cellular decomposition is not necessarily a triangulation and the geometry can be more discontinuous than a Regge geometry.

{\bf Spinnetworks (1971).}
In Loop Quantum Gravity, the \emph{spinnetwork} basis $\ket{\Gamma,j_\ell,\nu_n}$ \cite{Baez95a,Baez95aa, Rovelli:1995ac} is an orthonormal basis that diagonalizes the area and volume operators. The states in this basis are labelled by a graph $\Gamma$ and two quantum numbers coloring it: a spin $j$ at each link~$\ell$ and a volume eigenvalue~$\nu$ at each node~$n$.  The (dif\/f-invariant) Hilbert space $H_\Gamma$ obtained by considering only the states on the (abstract) graph $\Gamma$ is precisely the Hilbert space of an~$\SU(2)$ Yang--Mills theory on this lattice.   Penrose's ``spin-geometry'' theorem connects this Hilbert space with the description of the geometry of the cellular decomposition mentioned above: states in this Hilbert space admit a geometrical interpretation~\cite{Penrose:1971sn} as a quantum version of the 3-geometry (see~\cite{Rovelli:2011eq} and~\cite{Bianchi:2011zr}).  That is, a Regge 3-geometry def\/ined on a triangulation with dual graph $\Gamma$ can be approximated by semiclassical state in $H_\Gamma$.\footnote{Generically, the geo\-met\-ry def\/ined by the semiclassical states in~$H_\Gamma$ can be more general than a Regge geo\-met\-ry~\cite{Freidel:2010uq,Rovelli:2010km}. Furthermore, the graphs can be dual to generic cellular decomposition which are not triangulations, or to a cellular decomposition subjected to some specif\/ic restrictions~\cite{Hellmann:2011jn,Kaminski:2009fm,Kisielowski:2011vu}.}

{\bf Holonomies (1986).}
In the canonical quantization of General Relativity (GR), in order to implement Dirac quantization, it's convenient to choose the densitized inverse triad $E^{ia}$ (Ashtekar's electric f\/ield) and the Ashtekar--Barbero connection $A_a^i$  as conjugate variables \cite{Ashtekar86}, and then use the f\/lux of  $E^{ia}$ and the
the holonomy
$
{h}_\gamma= {\cal P}\exp [ \int_\gamma A_a^i ]
$,
namely the parallel transport operator for $A$ along a path $\gamma$, as fundamental variables for the quantization.
In the quantum theory, all relevant physical objects (partial obsevables \cite{Rovelli:2001bz}), for instance the operators for area and volume, have support only on these paths and their intersections. Considering a~truncation of the theory amounts to restricting the choice of the observables to a f\/inite subset.  In particular, the holonomies can be taken along the links of the graph, and the densitized inverse triad can be smeared over the faces of the triangulation. This connects the holonomy-triad variables to the discrete geometry picture.

\looseness=-1
The common  point of these dif\/ferent derivations is 3d coordinate gauge invariance, that has important consequences\footnote{The 3d coordinate gauge invariance
is the fundamental assumption of the LOST theorem  \cite{Fleischhack:2006zs, Lewandowski:2005jk}, that states the uniqueness of the representation in the LQG Hilbert space.}.
This invariance is the reason for the use of abstract graphs: it removes the physical meaning of the location of the graph on the manifold.
Therefore the graph we are considering is just a combinatorial object, that codes the adjacency of the nodes. Each node describe a quantum of space, and the graph describes the relations between dif\/ferent pieces of space. The Hilbert subspaces associated to distinct but topologically equivalent \emph{embedded} graphs are identif\/ied~\cite{Rovelli:1987df,Rovelli:1989za},  and each graph space ${h}_\Gamma$  contains the Hilbert spaces of all the subgraphs.

\begin{figure}[t]
\centering
\includegraphics[scale=0.3]{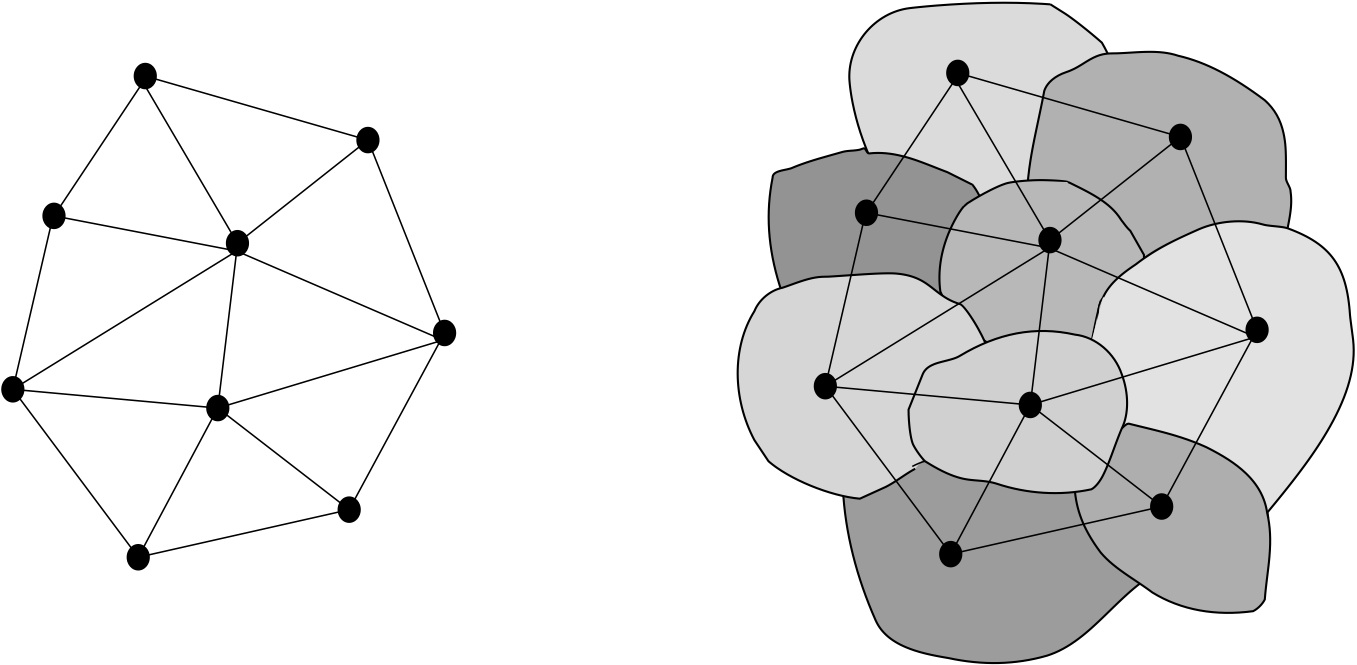}

$\hspace{-12em}|\Gamma, j_l,v_n\rangle$
\caption{To each node of a spinnetwork we can associate a ``quantum of space''.}
\label{ch1}
\end{figure}

\subsection{Doing physics with few nodes}\label{how}

Full general relativity is recovered from Regge calculus only by moving to triangulations with an arbitrary large number of cells; the full LQG Hilbert space includes arbitrary large graphs;  and the full set of observables cannot be restricted to holonomies and f\/luxes on a f\/ixed graph. The restriction to a f\/ixed triangulation or a f\/ixed graph amounts only to a \emph{truncation} of the theory, namely to disregarding an inf\/inite number of degrees of freedom and cutting down the theory to an approximate theory with a f\/inite number of degrees of freedom. But truncations are always needed in quantum f\/ield theory, in order to extract  numbers from the theory. For instance, numerical lattice QCD calculations are performed on a f\/inite lattice, and every given Feynman graph in QED involves only a f\/inite number of particles, hence a truncation of the full theory\footnote{It is interesting to remark that these two very dif\/ferent truncation schemes (f\/inite number of lattice sites in lattice QCD and f\/inite number of particles in perturbative QED) end up merging in quantum gravity. This is because a lattice site is a region of space and in quantum gravity a region of space is quantum of the gravitational f\/ield, very much like a particle is a quantum of a quantum f\/ield. Therefore both truncations are reinterpreted in quantum gravity as expansions in the number of a quanta: a f\/inite number of quanta of the gravitational f\/ield is described by a spin network, and is \emph{also} a lattice in the sense of QCD, as shown by the structure of $H_\Gamma$ indicated below. This convergence of the dif\/ferent quantum f\/ield theory pictures of lattice QCD and perturbative QED is one of the most beautiful aspects of LQG.}.  A given truncation can provide a valuable approximation to the full theory only under certain conditions and in certain physical regimes.  It is important to notice that in appropriate physical regimes even a low-order approximation can be ef\/fective. The Regge dynamics approximates the continuous GR dynamics when  $RL^2\ll 1$ where $R$ is the curvature scale and $L$ the length of the Regge bones\footnote{Namely when the Regge def\/icit angles, which code the curvature, are small.}. Therefore, as far as computing dynamics is concerned, a very rough triangulation can well approximate a near-f\/lat spacetime.  Very similarly, a f\/irst order Feynman diagram gives an excellent approximation to a scattering amplitude, even if the real spacetime trajectory of the particles is a smooth curved path, quite dif\/ferent from the piecewise straight path depicted in the Feynman diagram. Notice that since any given Feynman diagram involves a f\/inite number~$N$ of particle, the diagram is concretely def\/ined on the subspace~$H_N$ of Fock space spanned by the $n$-particle states with $n\le N$.

In LQG, the analog of $H_N$ is the state space ${H}_\Gamma$ formed by the (dif\/f-invariant) states on a~given graph is a subspace of the full Hilbert space ${H}_{\rm LQG}\sim \lim\limits_{\Gamma\to\infty} {H}_\Gamma$ \cite{Ashtekar:1993wf,Rovelli:2011eq}.   It has the structure
$
 {H}_\Gamma=  L^2 \big[\SU(2)^L/\SU(2)^n\big]$,
where $L$ the number of oriented links of $\Gamma$, $n$ is the number of nodes, and the group quotient is given by the gauge transformations at the nodes on the group elements on the links as in lattice gauge theory. It is therefore interesting to explore if we can compute physical quantities approximately working in ${H}_\Gamma$ instead of the full ${\cal  H}_{\rm LQG}$, in the same spirit of Quantum Field Theory  when one considers the Fock space for $N$ particles, instead of inf\/inite particles. This approach to LQG  is called \emph{graph expansion}, and relies on the invariance under dif\/feomorphism, that we have just discussed above.

Truncating the theory to a given f\/ixed $\Gamma$ corresponds to disregarding  the states that need a~``larger'' graph to be def\/ined, while all states that have support on graphs ``smaller'' than $\Gamma$ are already contained in  ${h}_\Gamma$.\footnote{States with support on graph smaller than $\Gamma$ (subgraphs of $\Gamma$) are already included in $H_\Gamma$.}

As the connection with Regge calculus shows, choosing a graph corresponds then to choosing an approximation for the system that we want to describe. Discretizing a continuous geometry by a given graph is nothing but coarse graning the theory. It is important to stress that the discreteness introduced by this process is dif\/ferent from the fundamental quantum discreteness of the theory. The f\/irst is the discreteness of the abstract graphs; the later is the discreteness of the spectra of the area and volume operator on each given~$H_\Gamma$. Mistaking these two sources of discreteness  has been a source of confusion in the literature\footnote{This is analog to the case of an electromagnetic f\/ield in a~box: the modes of the f\/ield are discrete and allow a~truncation of the theory, but quantum discreteness is something else: it is given by the quantized energy of each mode.}.

\begin{figure}[h]
  \centering
\includegraphics{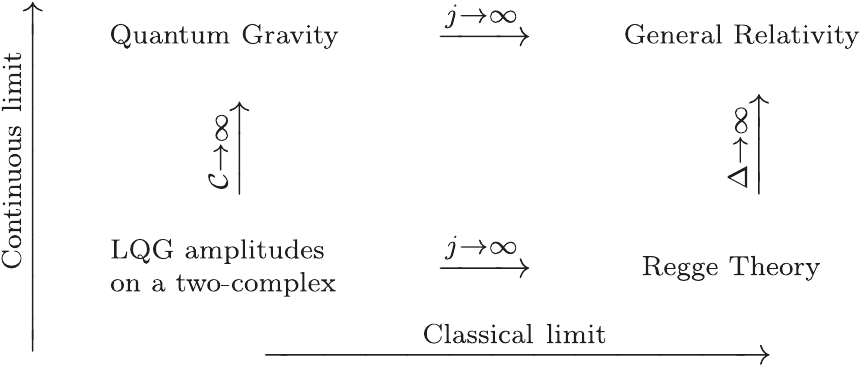}
  \caption{Continuous and classical limits in LQG.\label{limi}}
\end{figure}

Another source of confusion in the literature is the confusion between two dif\/ferent expansion: the graph expansion and the semiclassical expansion (see Fig.~\ref{limi}). The f\/irst is obtained by a~ref\/inement of the graph, the second by a large-distance limit~\cite{Rovelli:2011eq} on each graph. The f\/irst is an expansion valid in the regimes where a rough graph approximation is good (in particular at scales $L$ smaller that the curvature scale $R$); the second when we can disregard quantum ef\/fects,  (in particular, at scales $L$ larger that the Planck scale $L_P$).

This allows us to study two distinct limits in the theory:

{\bf Continuous limit.}
We are considering a discretized system with the further properties to be
 dif\/f-invariant system. This lead to a peculiarity while taking the continuum limit, as discussed in~\cite{Rovelli:2011uq}. In a theory like lattice QCD, this limit is achieved by sending the lattice spacing to zero and the coupling constant to its critical value. Here instead we just ref\/ine the graph, because the dynamics is not af\/fected by the size of the discretization since coordinates are unphysical.  This may have an important consequence: the discretization becomes nearly exact,  the number of nodes behaves as an ef\/fective expansion parameter and the system may approach a regime where the theory is topological (\emph{Ditt}-invariance regime~\cite{Rovelli:2011tt}). This might happen in describing homogeneous and isotropic geometries~\cite{Bahr:2009dz,Bahr:2009fv,Bahr:2010hc,Bahr:2011uj}.

{\bf Classical limit.} The fundamental discreteness of the theory is washed away when the  large-distance limit of the theory is taken. This corresponds to considering large spins, namely~$j\to\infty$. The classical theory is therefore recovered and the quantum parameters~$\hbar$ and~$\gamma$ (the Barbero--Immirzi parameter) disappear. Dif\/ferent ways to take the semi-classical limit, such that the quantum corrections would be under better control, are under study~\cite{Magliaro:2011fk}.

The classical continuous theory of General Relativity is recovered once both limits are taken. In quantum theory, these expansions  have provided interesting insights on the full theory. (This will be addressed further while addressing the covariant theory in Section~\ref{section4}.)

The point of view that we are presenting in this review is that we may not necessarily be obliged to deal with very complicated graph in quantum gravity. Interesting physics can arise even by considering a simple graph, with few nodes, and comparing our results with classical discrete gravity (Regge calculus).

Notice that, in Regge calculus, few nodes are already enough to capture the qualitative behavior of the model. This is  true for FLRW cosmologies. It has in fact been proven numerically~\cite{Collins:1973mh} that the dynamics of a closed universe, with homogeneous and isotropic geometry, can be capture by 5, 16 and 600 nodes (these are the regular triangulations of a 3-sphere) adapting the dynamics to these triangulations, and the only resulting dif\/ference is given by the scaling: having more tetrahedra, the growth is faster.

\begin{figure}[t]
\centering
\includegraphics[scale=0.28]{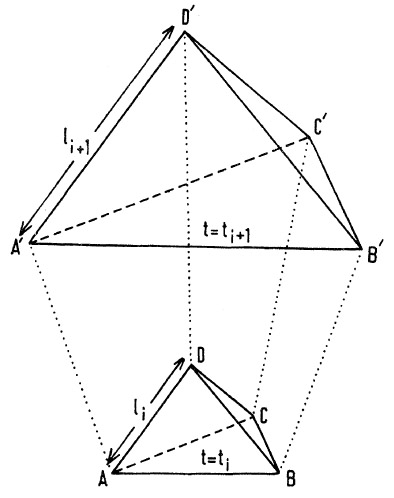}
\qquad
\includegraphics[scale=0.38]{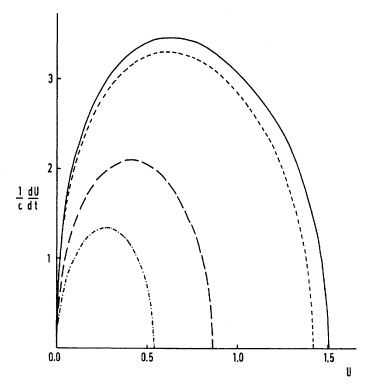}

\caption{On the left, a 4d building block of spacetime and, on the right, the evolution of 5, 16 and 500 of these building block (dashed lines), modeling a closed universe, compared whit the continuous analytic solution (solid line)~\cite{Collins:1973mh}. The qualitative behavior of these universes, coded in the rate of change of the volume, is the same.\label{Regge}}
\end{figure}

This review focuses on the construction of the theory when the graph is particularly small. In fact, we use here a minimal graph, given just by 2 nodes.

The main problem that we do \emph{not} address in this review is how radiative corrections can af\/fect the viability of the graph expansion.  In a renormalizable quantum f\/ield theory we know that radiative corrections can be split into a f\/inite part which is small, and therefore does not spoil the viability of the expansion, and a divergent part, which can be absorbed into a redef\/inition of f\/ields and coupling constants.  The same question arises in the present context: when ref\/ining the disctretization, do we generate corrections that are large and cannot be reabsorbed?  This question, upon which relies the viability of the entire philosophy of the graph expansion has not been  suf\/f\/iciently studied yet~\cite{Krajewski:2010yq, Perini:2008pd}.

\subsection{The cosmological interpretation}\label{modes}

The graph expansion can be put in correspondence with a mode expansion of the gravitational f\/ield on a compact space~\cite{Rovelli:2008ys}. (In the case of the dipole graph~-- see below~-- this correspondence has been worked out explicitly in~\cite{Battisti:2010kl}.) The truncation of the theory on a graph provides a~natural cut of\/f of the inf\/inite degrees of freedom of general relativity down to a f\/inite number.  Choosing a graph, we disregard  the higher modes of this expansion.
Therefore the truncation def\/ines an approximation viable for gravitational phenomena where the ratio between the largest and the smallest relevant wavelengths in the boundary state is small.

Notice that this is neither an ultraviolet nor an infrared truncation, because the whole physical space can still be large or small. What is lost are not wavelengths shorter than a given length, but rather wavelengths  $k$ times shorter than the full size of physical space, for some integer $k$.  To understand the nature of this approximation, we can refer again to numerical lattice QCD.
The number of lattice sites concretely needed for a numerical calculation is determined by the ratio between the smallest and largest wavelenghts involved in the phenomenon studied. For instance, for studying hadron masses~\cite{Durr:2008zz}, the relevant ratio is that determined by the hadron and quarks' sizes.

But the most striking example, where this kind of approximation applies, is given by cosmology itself.
Modern cosmology is based on the cosmological principle, that says that the dynamics of a homogeneous and isotropic space approximates well our universe. The presence of inhomogeneities (which in the real universe are large and well beyond perturbation theory at small scales) can be disregarded at a f\/irst order approximation, where we consider the dynamics as described \emph{at the scale} of the scale factor, namely the size of the universe. Thus our approximation depends on the scale factor, it is not just a large scale approximation: it depends on the ratio between the scale factor and the interaction that we want to consider. If we consider the dynamics of the whole universe, this ratio gives~1, and an unique degree of freedom is concerned. We can then recover the full theory adding degrees of freedom one by one. We obtain an approximate dynamics of the universe, with a f\/inite number of degrees of freedom. Postulating less symmetry, allows to add more degrees of freedom. So one can recover the full theory adding the degrees of freedom one by one. The specif\/ic choice of the truncation depends on the phenomena considered and the approximation desired.

In other words, working with a graph corresponds to choosing how many degrees of freedom we want to describe. A graph with a single degree of freedom is just one node: in a certain sense, this is the case of usual Loop Quantum Cosmology \cite{Ashtekar:2011ly}. To add degrees of freedom, we add nodes and links with a coloring. These further degrees of freedom
are a natural way to describe inhomogeneities and anisotropies~\cite{Battisti:2010kl, Rovelli:2008ys}, present in our universe.
When we ask  the graph to give a \emph{regular} cellular decomposition, node and links become indistinguishable, and we obtain back the unique FLRW degrees of freedom.

\begin{figure}[h]
\centering
\includegraphics{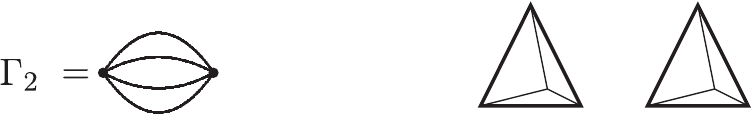}
\caption{The ``dipole'' graph $\Gamma_2$ is given by 2 nodes and 4 links. It is dual to a triangulation of a~3-sphere, where 2 tetrahedra are glued together by identifying their faces.\label{dipole}}
\end{figure}

The easiest thing that can be done is to pass from $n=1$ to $n=2$ nodes. We choose to connect them by $L=4$ links, because in this way the dual graph will be two tetrahedra glued together, and this can be viewed as the triangulation of a 3-sphere (Fig.~\ref{dipole}). Note that we are not obliged to chose a graph corresponding to a triangulation, but this turn out to be very useful when we want to associate an intuitive interpretation to our model.
In order to understand how this can be concretely use to do quantum gravity and quantum cosmology, we need to place on the graph the $\SU(2)$ variables.

\section{The Hamiltonian dipole}\label{section2}

In this section we review the original dipole construction, and the relation between its variables and those of the cosmological models.

\subsection{LQG phase space}\label{phsp}

Loop Quantum Gravity is a general covariant quantum f\/ield theory, where the 3d coordinate gauge invariance is ref\/lected in the choice of $\SU(2)$ as fundamental group. Let us start by associating the group element ${h}_\ell\in \SU(2)$ and a $\su(2)$ algebra element $E_\ell=E_\ell^i\tau_i$\, where $\{\tau_i \,|\  i=1,2,3\}$ is a basis in $\su(2)$  to the links $\ell$ of a given graph $\Gamma$.

The cotangent boundle of $\SU(2)^{L}$ and its natural symplectic structure give the phase space of the theory: ${h}_\ell$ and $E_\ell$ are phase space variables with the conventional Poisson
brackets structure of a canonical lattice $\SU(2)$ Yang--Mills theory, that is\footnote{Here we have assumed $\ell'\ne \ell^{-1}$. If $\ell'=\ell^{-1}$, the Poisson brackets are obtained using the equations ${h}_{\ell^{-1}}=U^{-1}_\ell
$ and the algebra element $E_{\ell^{-1}}= -  {h}_{\ell}^{-1}E_{\ell}{h}_{\ell}$.}
\begin{gather}
\{  {h}_\ell,{h}_{\ell'} \} = 0  , \qquad
\{  E_\ell^i,{h}_{\ell'} \} =  \delta_{\ell\ell'}  \tau^i {h}_{\ell}   , \qquad
\{  E_\ell^i,E_{\ell'}^j \} = -    \delta_{\ell\ell'}  \epsilon^{ijk} E_\ell^k   .\label{tree}
\end{gather}
As in lattice QCD, a quantum representation of the observable algebra~\eqref{tree} is provided by the Hilbert space ${\mathcal H}_{\rm aux}=L^2[\SU(2)^{L}, \dd {h}_\ell]$ where $\dd {h}_\ell$ is the Haar measure.  The operators~${h}_\ell$ are diagonal and the operators $E_{\ell}$ are the left invariant vector f\/ields
on each $\SU(2)$. The operators~$E_{\ell^{-1}}$ turn then out to be the right invariant vector f\/ields.  These operators satisfy\footnote{Notice that these variables transform properly under internal gauge transformations. In fact we can  write $G[\lambda]:=2\sum_n\tr[\lambda_ n G_n]$ where $\lambda_n\in su(2)$, the inf\/initesimal gauge transformation of ${h}_{\ell}$ is $\delta {h}_{\ell}=\{{h}_{\ell},G[\lambda] \}=\lambda_{n_1}{h}_{\ell}-{h}_{\ell}\lambda_{n_2}$ where $(n_1, n_2)$ are respectively the source and the target of of the link $\ell$.}
 the
\begin{gather}\label{gauge}
\mbox{gauge constraint}\qquad
  G_n \equiv  \sum_{\ell} E_{\ell} \sim 0
  \qquad
   \forall\, \ell\in n,
\end{gather}
that can be seen as a closure condition on the cell dual to the node~$n$.

The {\it states} that solve the gauge constraint~(\ref{gauge}) are labeled by $\SU(2)$ spinnetworks on the graph $\Gamma$.    A basis of these is given by states $|j_{\ell}, \iota_n\rangle$, where $\ell=1,\dots ,L$ and $n=1,\dots ,N$ range over the links and the nodes of the graph. These are def\/ined by
\begin{gather*}
\mbox{spinnetwork states}\qquad
 \psi_{j_{\ell}\iota_n}({h}_{\ell}) \equiv\langle {h}_{\ell} | j_{\ell}, \iota_n \rangle \equiv  \otimes_{\ell}\;  \Pi^{(j_{\ell})}({h}_{\ell})  \cdot  \otimes_n \; \iota_n,
\end{gather*}
where $\Pi^{(j)}(U)$ are the matrix elements of the spin-$j$ representation of $\SU(2)$ and ``$\cdot$'' indicates the contraction of the indices of these matrices with the indices of the intertwiners~$\iota_n$ dictated by the graph $\Gamma$. For details, see~\cite{Tate,Rovelli,Thiemann}.

Operators for area and volume can be constructed in terms of $E_{\ell}$. For each link $\ell$ we can associate the area of the face punctured by the link in the dual cellular decompositon
\begin{gather} \mbox{{area}}\qquad
 {A}_\ell=\sqrt{ {E}_\ell
  {E}_\ell}=8\pi \gamma \lp^2 \sqrt{j_\ell(j_\ell+1)},
\label{area}
\end{gather}
where $\gamma$ is the Barbero--Immirzi parameter,
and for each node we can associate the volume of the cell on which the node $n$ is sitting. The expression for a generic $n$-valent node is available but complicate~\cite{Bianchi:2011zr}, while in the simplest 4-valent case the expression becomes just
\begin{gather*}
\mbox{{volume$^2$}}\qquad
V_n^2=   \frac14 \sum_{\ell\ell'\ell''\in n} \tr[E_{\ell}E_{\ell'}E_{\ell''}]=
 \tr[E_{\ell}E_{\ell'}E_{\ell''}].
\end{gather*}
where we have chosen the links $\{\ell,\ell',\ell''\}$ to have positive orientation. Notice that the sum over the four unordered triplets of \emph{distinct} links drops because of~\eqref{gauge}.
The total volume will be just  ${V}_\Gamma=\sum {V}_n$ $\forall\, n\subset \Gamma$.
The dynamics is governed by the Hamiltonian constraint. Recall that the Hamiltonian constraint for general relativity can be written as~\cite{Thiemann}:
\begin{gather*} 
C_{\mathrm{grav}} = \int_{S^3} \dd^3 x\,
\epsilon^{ij}{}_{k}  {E^{ai}E^{bj}}  F_{ab}^k  -
2\big(1+\gamma^2\big)
 \int_{S^3} \dd^3 x\,
  {E^{ai}E^{bj}} K_{[a}^i K_{b]}^j
:= C + C_\gamma,
\end{gather*}
where $K_a^i $
is the extrinsic curvature, and where we have set the lapse equal to the total 3-volume.  Let us concentrate for the moment on the f\/irst term.
If the discretization is appropriately chose before the quantization, we can write the f\/irst term of the Hamiltonian constraint by approxmating the f\/ield-strength tensor $F_{ab}^k$ with the holomy $h_{\ell\ell'}$  such that
\begin{gather}\label{ham}  
C=\sum_n C_n \qquad \mbox{with} \qquad
C_n =  \sum_{\ell\ell'\in n} \tr[
{h}_{\ell\ell'}E_{\ell'}E_{\ell}
] \sim 0.
\end{gather}
This form was suggested in the early day of LQG \cite{Rovelli:1993bm} and is nowadays exploited in loop cosmology~\cite{Ashtekar:2008pi, Ashtekar:2011ly}.

To complete our construction, at each node we couple a scalar f\/ield
$\phi_n$, with conjugate momentum $p_{\phi_n}$,
that work as
a family of multif\/ingered ``clock'' variables \cite{Ashtekar:2007tv, Ashtekar:2003hd,Ashtekar:2008pi,Ashtekar:2006rx}. We need to introduce this physical clock because otherwise we would not be able to  keep track of evolution in a background-independent manner~\cite{Dittrich:2005kc,Dittrich:2004cb,Dittrich:2007jx,Rovelli:2001bz,Rovelli:1990pi, Rovelli:1990ph}.
The introduction of a scalar f\/ield provides also  a simplif\/ied manner to model the matter content of the universe.  The generalization to a more realistic description is straightforward: here we will choose an ultra-local scalar f\/ield, but one can  add the spacial derivative terms in the Hamiltonian of the matter f\/ield, and the description of Yang--Mills and fermion f\/ields is particularly well-adapted to this formalism~\cite{Tate,Rovelli,Thiemann}.

Therefore the total Hamiltonian constraint is
\begin{gather}\label{hamot}
C_{\rm tot}=
\sum_n \sum_{\ell\ell'\in n} \left(\tr[
{h}_{ \ell \ell'}
 E_{\ell'}E_{\ell}
 ]  +C_\gamma+ 4\pi G  p_{\phi_n}^2  \right)\sim 0,
\end{gather}
where $G$ is the Newton constant, determining the matter-gravity coupling.
With a scalar f\/ield, the Hilbert space becomes ${\mathcal H}_{\rm aux}=L_2[\SU(2)^{L}, d{h}_{\ell}]\otimes L_2[R^n]$, with a (generalized) basis $ | j_{\ell}, \nu_n,\phi_n \rangle $ and the states can be written in the form
\begin{gather*}
\psi(j_{\ell}, \nu_n,\phi_n ) \equiv \langle  j_{\ell}, \nu_n, \phi_n  | \psi \rangle.
\end{gather*}
In this basis the operator $\phi_n$ is diagonal while $p_{\phi_n}=-i\frac{\partial}{\partial \phi_n}$.

\subsection{Dipole cosmology}

Consider the simple case obtained by taking $n=2$ and the natural triangulation of the a three-sphere $S^3$ obtained by gluing two tetrahedra by all their faces, as in Fig.~\ref{dipole}. This represents a~f\/inite dimensional truncation of LQG and describes the Bianchi IX Universe plus six inhomogeneous degrees of freedom \cite{Battisti:2010kl}. The gravitational variables are $({h}_{\ell}, E_{\ell})$,  $\ell=1,2,3,4$.  We have two Hamiltonian constraints, whose algebra is naturally closed given the simplicity of the system considered. The $\SU(2)$ symmetry structure enters twice in our description: not only in the discretization of the Ashtekar--Barbero variables, but also in order to add the inhomogeneities.

The gravitational Hilbert space is $L^2[\SU(2)^4/\SU(2)^2]$ and a basis of spinnetwork states that solve the gauge constraint is given by the states $|j_\ell, \iota_n\rangle=|j_1,j_2,j_3,j_4, \iota_1, \iota_2\rangle$.  The action of one gravitational Hamiltonian constraint on a state gives
\begin{gather*}
\tilde C |j_\ell, \iota_n\rangle = \sum_{\ell\ell'} C_{\ell\ell'}|j_\ell, \iota_n\rangle,
\end{gather*}
where  each term of the sum comes from one of the terms in the sum in $\ell$ and $\ell'$ in~(\ref{ham}).
More explicitly, we have
\begin{gather*}
C_{12}|j_1,j_2,j_3,j_4, \iota_1, \iota_2\rangle
=\sum_{\epsilon,\delta=\pm 1}
C^{\epsilon\delta \iota'_1\iota'_2 }_{j_f\iota_1\iota_2}   |j_1+\frac\epsilon{2},j_2+\frac\delta{2},j_3,j_4, \iota'_1, \iota'_2\rangle,
\end{gather*}
because the operator $U_{12}=U_1U_2^{-1}$ in (\ref{ham}) multiplies the terms $\Pi^{j_1}(U_1)$ and $\Pi^{j_2}(U_2)$ and
\begin{gather*}
U\Pi^{j}(U)= \Pi^{1/2}(U)\Pi^{j}(U)= c_+\Pi^{j+1/2}(U)+c_-\Pi^{j-1/2}(U).
\end{gather*}
The matrix elements $C^{\epsilon\delta \iota'_1\iota'_2 }_{j_\ell\iota_1\iota_2} $ can be computed with a straightforward exercise in recoupling theory from~\eqref{ham} (and from the Hamiltonian that makes use of the ``Thiemann's trick'' with some more algebra).  In a dif\/ferent notation, in terms of the wave function components, we can write
\begin{gather*}
\tilde C \psi(j_{\ell}, \iota_n)
=\sum_{\epsilon_j=0,\pm 1}
{C}^{\epsilon_{\ell} \iota_n'}_{j_{\ell} \iota_n}  \psi\left(j_{\ell}+\frac{\epsilon_{\ell}}{2}, \iota'_n\right),
\end{gather*}
where $C^{\epsilon_j \iota_n'}_{j_{\ell} \iota_n}$ vanishes unless $\epsilon_{\ell}=0$ for two and only two of the four $j$'s.
The scalar f\/ield variables are~$\phi_1$,~$\phi_2$. Taking these into account leads to the wave functions $\psi(j_{\ell}, \iota_n, \phi_n)$, and (\ref{hamot}) gives the dynamical equations
\begin{gather} \label{h1}
\left( \frac{\partial^2}{\partial \phi_1^2}
 + \frac{\partial^2}{\partial \phi_2^2}\right)\psi(j_{\ell}, \iota_n, \phi_n) =
\frac2\kappa \sum_{\epsilon_{\ell}=0,\pm 1}
C^{\epsilon_{\ell} \iota_n'}_{j_{\ell} \iota_n}\  \psi\left(j_{\ell}+\frac{\epsilon_j}{2}, \iota'_n,\phi_n\right),\\
\label{h2}
\frac{\partial^2}{\partial \phi_1^2}\psi(j_{\ell}, \iota_n, \phi_n) =
\frac{\partial^2}{\partial \phi_2^2}\psi(j_{\ell}, \iota_n, \phi_n).
\end{gather}
The coef\/f\/icients $C$ can be computed explicitly from recoupling theory.  They vanish unless two~$\epsilon_{\ell}$'s are zero.  Equations (\ref{h1}), (\ref{h2}), def\/ined on Hilbert space ${\mathcal H}_2=L_2[\SU(2)^4/\SU(2)^2]\otimes L_2[R^2]$ def\/ine a quantum cosmological model which is just one step out of homogeneity.

\subsubsection[Born-Oppenheimer approximation and LQC]{Born--Oppenheimer approximation and LQC}

We now ask if and how LQC is contained in the model def\/ined above.
The state space $H_2$ contains a subspace that could be identif\/ied as a homogeneous universe.  This is the subspace $H^{\rm hom}\subset H_2$ spanned by the states $|j,j,j,j, \iota_j,\iota_j,\phi,\phi\rangle$ where $\iota_j$ is the eigenstate of the volume that better approximates the volume of a classical tetrahedron whose triangles have area $j$.  However, the dynamical equations (\ref{h1}), (\ref{h2}) do not preserve this subspace. This is physically correct, because the inhomogeneous degrees of freedom cannot remain sharply vanishing in quantum mechanics, due to Heisenberg uncertainty.  Therefore it would be wrong to search for states that reproduce LQG \emph{exactly}, within this model.   In which sense then can a quantum homogeneous cosmology make sense?

The answer should be clear thinking to the meaning of the cosmological principle, that is at the base of every cosmological model.   The cosmological principle is the hypothesis that in the theory there is a regime where the inhomogeneous degrees of freedom do not af\/fect too much the dynamics of the homogeneous degrees of freedom, and that the state of the universe happens to be within such a regime. In other words, the homogeneous degrees of freedom can be treated as ``heavy'' degrees of freedom, in the sense of the Born--Oppenheimer approximation, and the inhomogeneous one can be treated as ``light'' ones. Let us therefore separate explicitly the two sets of degrees of freedom.
This can be done as follows.

First, change variables from the group variables $h_{\ell}\in \SU(2)$ to algebra variables $A_{\ell}\in \su(2)$,  def\/ined by $\exp A_{\ell}=h_{\ell}$.\footnote{This is only a convenient rewriting of the holonomies, not really a return of the connection as main variable.}  Following  what is done in Loop Quantum Cosmology~\cite{Ashtekar:2003hd},  let us f\/ix a~f\/iducial $\su(2)$ element $\omega_{\ell}\in \su(2)$ for each link $\ell$.  We choose for simplicity a f\/iducial connection normalized as $|\omega_{\ell}|=1$, and
such that the four vectors $\omega_{\ell}$ are normal to the faces of a regular tetrahedron centered at the origin of $\su(2)\sim R^3$.
Our variables can be decomposed into
\begin{gather}
A_{\ell}=c \omega_{\ell}+a_{\ell}, \qquad
E_{\ell}=p \omega_{\ell}+e_{\ell}.  \label{din}
\end{gather}
In order to f\/ix this decomposition uniquely, we impose the following conditions: $p$ has to be determined by the total volume
$V= p^{3/2}$,
and $c$ should be its conjugate variable so that
\begin{gather*}
\{c,p\} = \frac83\pi\gamma G.
\end{gather*}
The variable $c$ can then be identif\/ied at the classical level with the scalar coef\/f\/icients multiplying respectively the extrinsic and intrinsic curvature, namely we have $c\sim\gamma\dot a +1$ as in LGC, where $\dot a$ is the time derivative of the scale factor. We also def\/ine $\Delta V=V_2-V_1$, so that $V_{1,2}=\frac12(V\pm\Delta V)$.
In the quantum theory, $E_\ell$ turns  out to be a  left invariant vector f\/ield, call it $L_\ell$, so that~\eqref{din} yields the decomposition
\begin{gather*}
L_\ell= \omega_{\ell} \frac{\partial}{\partial c} +   \tilde L_\ell ,
\end{gather*}
where $\tilde L_\ell c =0$.
Inserting this decomposition into the Hamiltonian constraint (\ref{ham}) gives
\begin{gather*}
\tilde{C_n} =\sum_{\ell\ell'\in t} \tr\left[
e^{c\omega_{\ell}-a_{\ell}}e^{-c\omega_{\ell'}-a_{\ell'}}
\left( \omega_{\ell'} \frac{\partial}{\partial c} +   \tilde L_\ell \right)
\left(\omega_{\ell} \frac{\partial}{\partial c} +   \tilde L_{\ell'}\right)\right].
\end{gather*}

Let us now decompose this constraint into two parts, the f\/irst of which depends only on the homogeneous variable~$c$.  This can be done keeping only the f\/irst term of the expansion of the exponentials in $a_{\ell}$ and $a_{\ell'}$, and only the $V$ term in the volume term. That is, we write
\begin{gather*}
C_n =\frac{1}{2}C^{\rm hom}+C_n^{\rm inh},
\end{gather*}
where
\begin{gather}\label{hhom}
\tilde{C_n} =\sum_{\ell\ell'\in t} \tr\left[
e^{c\omega_{\ell}}e^{-c\omega_{\ell'}}
\left( \omega_{\ell'} \frac{\partial}{\partial c} \right)
\left(\omega_{\ell} \frac{\partial}{\partial c} \right)\right].
\end{gather}
The interpretation of this spilt is transparent: $C^{\rm hom}$ gives the gravitational energy in the homogeneous degree of freedom, while  $C_n^{\rm inh}$ gives the sum of the energy in the
inhomogeneous degrees of freedom and the interaction energy between the two sets of degrees of freedom.  Finally, we write the homogeneous variable $\phi=\phi_1+\phi_2$ and $\phi_-=\phi_1-\phi_2$.

Following Born and Oppenheimer, let us now make the hypothesis that the state can be rewritten in the form
\begin{gather}\label{bh}
\psi(U_{\ell},\phi_n)=\psi_{\rm hom}(c,\phi) \psi_{\rm inh}(c,\phi;a_{\ell},\phi_-),
\end{gather}
where the variation of $\psi_{\rm inh}$ with respect to $c$ and $\phi$ can be neglected at f\/irst order. Here $\psi_{\rm hom}$  represents the quantum state of the homogeneous cosmological variables, while $\psi_{\rm inh}$ represents the quantum state of the inhomogeneous f\/luctuations over the homogeneous background $(c,\phi)$.
Inserting the Born--Oppenheimer ansatz~(\ref{bh}) into the Hamiltonian constraint equation, and taking $N_1=N_2$, we have the equation
\begin{gather*} 
\frac{\kappa}2  \psi_{\rm inh}
\frac{\partial^2}{\partial \phi^2}\psi_{\rm hom}
+\frac{\kappa}2   \psi_{\rm hom}
\frac{\partial^2}{\partial \phi_-^2}\psi_{\rm inh}
-\psi_{\rm inh} \tilde C^{\rm hom}\psi_{\rm hom}-  \tilde C^{\rm inh}\psi_{\rm hom}\psi_{\rm inh}=0.
\end{gather*}
Dividing by $\psi_{\rm hom}\psi_{\rm inh}$ this gives
\begin{gather*}
\frac{\frac{\kappa}2  \frac{\partial^2}{\partial \phi^2}\psi_{\rm hom}}{\psi_{\rm hom}} -
\frac{  \tilde C^{\rm hom}\psi_{\rm hom}}{\psi_{\rm hom}}
 =
-\frac{\frac{\kappa}2  \frac{\partial^2}{\partial \phi_-^2}\psi_{\rm inh}}{\psi_{\rm inh}} +
\frac{  \tilde C^{\rm inh}\psi_{\rm hom}\psi_{\rm inh}}{\psi_{\rm hom}\psi_{\rm inh}}.
\end{gather*}
Since the left hand side of this equation does not depend on the inhomogeneous variables,
there must be a function $\rho(c,\phi)$ such that
\begin{gather*}
  \frac{\kappa}2 \frac{\partial^2}{\partial \phi^2}\psi_{\rm hom} -
  \tilde C^{\rm hom}\psi_{\rm hom}- \rho{\psi_{\rm hom}}=0, \\
  \frac{\kappa}2 \frac{\partial^2}{\partial \phi_-^2}\psi_{\rm inh} +
\frac{ \tilde C^{\rm inh}\psi_{\rm hom}\psi_{\rm inh}}{\psi_{\rm hom}}=\rho{\psi_{\rm inh}}.
\end{gather*}
The second equation is the Schr\"odinger equation for the inhomogeneous modes in the background homogeneous cosmology $(c,\phi)$, where $\rho(c,\phi)$ plays the role of energy eigenvalue. The f\/irst equation is the quantum Friedmann equation for the homogeneous degrees of freedom $(c,\phi)$, corrected by the energy density $\rho(c,\phi)$ of the inhomogeneous modes.  At the order zero of the approximation, where we disregard entirely the ef\/fect of the inhomogeneous modes on the homogeneous modes, we obtain
\begin{gather*}
\frac{\kappa}2  \frac{\partial^2}{\partial \phi^2}\psi_{\rm hom}
=   \tilde C^{\rm hom}\psi_{\rm hom}.
\end{gather*}
Let us now analyze the action of the operator $C^{\rm hom}$, def\/ined in (\ref{hhom}). Notice that $c$ multiplies the generator of a $\U(1)$ subgroup of $\SU(2)^4$. Therefore it is a periodic variable $c\in[0,4\pi]$.  We can therefore expand
the states $\psi_{\rm hom}(c,\phi)$ in Fourier sum
\begin{gather*}
\psi_{\rm hom}(c,\phi) = \sum_\mu \psi(\mu,\phi)  e^{i\mu c/2},
\end{gather*}
where $\mu$ is an integer.  The basis of states  $\langle c\,|\mu\rangle=e^{i\mu c/2}$ in the gravitational sector of the $\psi_{\rm hom}$'s state space satisf\/ies
\begin{gather*}
p^{\frac32}|\mu\rangle = k \mu^{\frac32}|\mu\rangle,\qquad
-4\sin^2(c/2)|\mu\rangle = |\mu+2\rangle-2|\mu\rangle+|\mu-2\rangle,
\end{gather*}
which we shall use below\footnote{The explicit relation between these states and the states in the inhomogeneous-model state-space is not straightforward, and will be investigated in detail  elsewhere.}. Here $k=\left(\frac{8\pi G\gamma}6\right)^{\frac32}$.

The homogeneous Hamiltonian constraint (\ref{hhom}) can be rewritten as
\begin{gather*}
\tilde{C_t}^{\rm hom} =\sum_{\ell\ell'\in t} \tr[e^{c\omega_{\ell}}e^{-c\omega_{\ell'}}
\omega_{\ell'} \omega_{\ell}] \frac{\partial}{\partial c}
 \frac{\partial}{\partial c}   \equiv        \frac12 C^{\rm hom} .
\end{gather*}
This can be rewritten as \cite{Battisti:2010kl}
\begin{gather*}
\tilde{C}^{\rm hom} = \sum_{\ell\ell'} \!\tr\!\left[
\left(\cos{\frac c2}1\!\!1+2\sin{\frac c2}\omega_{\ell}\right)\!
\left(\cos{\frac c2}1\!\!1-2\sin{\frac c2}\omega_{\ell'}\right)
\omega_{\ell'} \omega_{\ell}\right] \frac{\partial^2}{\partial c^2}
=  \f{17}6 (\cos c -1)\frac{\partial^2}{\partial c^2}.
\end{gather*}
The action of this operator on the states $\psi_{\rm hom}(\mu,\phi)$ is therefore easily computed
\begin{gather*}
\tilde{C}^{\rm hom}\psi_{\rm hom}(v,\phi)= \f{17}6 [
\mu^2 \psi_{\rm hom}(\mu+2,\phi)
-\mu^2 \psi_{\rm hom}(\mu,\phi)
\mu^2 \psi_{\rm hom}(\mu-2,\phi)].
\end{gather*}
Bringing everything together, the full equation (\ref{hhom}) reads{\samepage
\begin{gather}
C^+(\mu)  \psi_{\rm hom}(\mu+2,\phi)+C^0(\mu)  \psi_{\rm hom}(\mu,\phi)+C^-(\mu) \psi_{\rm hom}(\mu-2,\phi)\nonumber\\
\qquad{}+ \frac{\partial^2}{\partial \phi^2}\psi_{\rm hom}(\mu,\phi) =0,\label{din2}
\end{gather}
where the coef\/f\/icient take the simple form $C^\pm(\mu)= - \frac12 C^0(\mu)=\frac{\mu^2}{2\kappa} $.}

Equation~\eqref{din2} has the structure of the LQC dynamical equation.  Thus, a structure very similar to the one of LQC appears in the zero order Born--Oppenheimer approximation of a Loop Quantum Gravity quantization of a f\/inite number of degrees of freedom of the gravitational f\/ield, truncated according to the approximation dictated by the cosmological principle.

We stress that we are not claiming here that the precise form of LQC is recovered. In particular, the two Hamiltonian operators might dif\/fer, because of possible dif\/ferent values of the coef\/f\/icients. The exact relation between the two theories, and the possibility of adapting the quantization scheme to the precise form of LQC have not yet been investigated in detail.

{\bf Bounce.}
It is easy to see that, given the periodicity in $c$ in the gravitational part of the Hamiltonian constraint, the model we have presented leads to a bounce. A f\/irst study of this has been presented in \cite{Battisti:2010uq}. In fact, the total Hamiltonian for the system reads
\begin{gather*}
\mathcal H=\mathcal{H}_{g}+\mathcal{H}_{m}=\frac{17}{6} p^2  ( \cos(c-\alpha) -1 ) +|p|^{3}\rho ,
\end{gather*}
where $\rho=\rho(t)$ is the matter energy density. Studying the equations of motion we obtain a~modif\/ied Friedmann equation of the form
\begin{gather*}
\left(\f{\dot a}a\right)^2=\left( \frac{\dot{p}}{2 p} \right)^2\propto \rho\left(1-\f\rho{\rho_c}\right) .
\end{gather*}
Here $\rho_c$, playing the role of the critical density at which the bounce occurs, is def\/ined as $\rho_c=17/(3p)$ \cite{Battisti:2010uq}. We notice that such a density depends on $p$, giving rise to incompatibility with a~proper classical limit of the model~\cite{Corichi:2008fu}. This is a situation already found in the old models of LQC and we know it depends on a naive regularization of the Hamiltonian constraint. In LQC there is a preferred regularization, that corresponds with the so-called $\bar\mu$-scheme~\cite{Ashtekar:2006wn}: the critical density is a constant in this way, and its value depends on the minimal are gap provided by LQG, i.e.\ the minimal eigenvalue given by~\eqref{area}. At the moment we write, the implementation of this kind of regularization in the dipole model is under study.  We refer the interested reader to~\cite{Livine:2011up}, in particular for what concerns the implementation of the $\bar\mu$-scheme in the spinor framework, that will be introduced in Section~\ref{section3}.

\subsubsection{Anisotropies and inhomogeneities} \label{anis}

In the previous treatment we have performed the Bohr--Oppenheimer approximation in such a way to extract the heavier degree of freedom, given by the homogeneous and isotropic one coded in the scale factor.
On the other hand, we have already mentioned that, using the dipole graph as the base of our model, we are in a natural situation to accommodate more that a single degree of freedom. In fact, the Hilbert space of the dipole graph contains 6 degrees of freedom. These should code information about the presence of anisotropy and inhomogeneity, but haw can we relate them to a classical interpretation? In classical General Relativity, the maximal amount of anisotropy that a system with the topology of a 3-sphere can have, is described by the Bianchi IX model. This is usually charachterized
 by three scale factors $a_I=a_I(t)$, which identify three independent directions in the time evolution of the Cauchy surfaces.  In the connection formalism, we have to consider three dif\/ferent connections $c_I=c_I(t)$ and momenta $p_I=p_I(t)$. Therefore the basic variables \eqref{din} in our model takes the form
\begin{gather}\label{varbix}
h_\ell=\exp\left(c^I\omega^I_\ell\tau_I\right), \qquad E_\ell=p^I\omega^I_\ell\tau_I.
\end{gather}
Notice that here the holonomies \eqref{varbix} are simply group elements taken on the links of the dipole, without any specif\/ic orientation. The connection components are summed over and they are thus independent on the $I$-direction. This dif\/fers from the situation in the LQC Bianchi IX model \cite{WilsonEwing:2010rh}, where  the basic holonomies $h_I$ are computed along paths parallel to the three axis individuated by the anisotropies.

The anisotropies of the Bianchi IX model requires only three degrees of freedom. The re\-maining large-scale gravitational degrees of freedom captured by the dipole dynamics are necessarily  inhomogeneous. Recall that in Section~\ref{modes} we introduced the idea that these degrees of freedom can be added one by one as the terms of an expansion in modes. Therefore we consider here an expansion of the gravitational f\/ields in tensor harmonics: the lowest modes should be the ones captured by the dipole dynamics. Luckily a similar mode expansion around Bianchi IX has been already studied by Regge and Hu using Wigner functions~\cite{Hu:1972zz}. The Wigner func\-tions~$D^j_{\alpha' \alpha}(g(x))$ determine a basis of functions of the symmetry-group of the model. Recall that we can use group elements $g(x)$ to coordinatize the physical space that has the $S^3$ topology.

Let us adapt this formalism to the f\/irst order variables we use.
We start with the triads. We write a generic perturbed triad
${E}_{I}^a(x,t)$ as the sum of the background triad $e_I$
f\/ield and a~perturbation
\begin{gather*} 
{E}_{I}^a(x,t)={e}_{I}^a(x)+ \psi_{IJ}(x,t)  e_{J}^a(x) .
\end{gather*}
We write this as a sum of
components of def\/inite $j$ and $\alpha$ quantum numbers
\begin{gather*}
\psi_{IJ}(x,t)= \sum_{ j \alpha}  \psi^{ j \alpha}_{IJ}(x,t)  ,
\qquad \mbox{where}\qquad
\psi^{ j \alpha}_{IJ}(x,t)
=
 \sum_{ \alpha'=-j}^j   \psi_{IJ}^{j\alpha\alpha'}(t)    D^j_{\alpha' \alpha}\left(g(x)\right) .
\end{gather*}
The same can be done for the connection
\begin{gather*} 
\omega^{I}_a(x) \rightarrow  \tilde{\omega}^{I}_a(x,t)=\omega^{I}_a(x)
+ \varphi^{IJ}(x,t)  \omega^{J}_a(x) .
\end{gather*}
Expanding this in  components of def\/inite $j$ and $\alpha$ quantum numbers gives
\begin{gather*}
\varphi^{IJ}(x,t)= \sum_{ j \alpha}  \varphi_{ j \alpha}^{IJ}(x,t)  ,
\qquad \mbox{where}\qquad
\varphi_{ j \alpha}^{IJ}(x,t)
=
 \sum_{ \alpha'=-j}^j   \varphi^{IJ}_{j\alpha\alpha'}(t)    D^j_{\alpha' \alpha}\left(g(x)\right) .
\end{gather*}
The $(\varphi^{IJ}_{j\alpha\alpha'}(t),\psi^{IJ}_{j\alpha\alpha'}(t))$ are the time-dependent expansion coef\/f\/icients that capture the inhomogeneous degrees of freedom. They, are given by matrices in the internal indices~$I$,~$J$, labeled by the spin~$j$ that runs from~$j=1/2$ to all the semi-integers numbers, and the corresponding magnetic number~$\alpha$.

We want to have the inhomogeneities determined by namely nine degrees of freedom coded in~$\varphi^{I}_{\alpha}(t)$,~$\psi^{I}_{\alpha}(t)$. This is achieved by assuming that the matrices $(\varphi^{IJ}_{j\alpha\alpha'}(t),\psi^{IJ}_{j\alpha\alpha'}(t))$ are diagonal in the internal indices $I$,~$J$ and it is dif\/ferent from zero only for lowest nontrivial integer spin $j=1$ and for, say, $\alpha=0$. That is, we restrict to the components
\begin{gather*}
 \varphi^{IJ}_{1,0,\alpha}(t)=\delta^{IJ}\varphi^{I}_{\alpha}(t), \qquad
 \psi^{IJ}_{1,0, \alpha}(t)=\delta^{IJ}\psi^{I}_{\alpha}(t), \qquad
 \alpha=-1,0,1.
\end{gather*}

The Gauss constraint reduces further the degrees of freedom to six, which is the number of degrees of freedom captured by the dipole
variables. Therefore, we can interpret the six extra degrees of freedom of the dipole model (beyond anisotropies), as a description of the diagonal part of the lowest integer mode of the inhomogeneities.
In this way the variables of the dipole model can be connected to the quantities  $(\varphi^{I}_{\alpha}(t), \psi^{I}_{\alpha}(t))$.

In order to complete the connection with the dipole variables, let us consider the f\/iducial-algebra elements $\omega_f^I$. these are perturbed as well, and as a consequence the 1-forms $\tilde\omega^I$ no longer satisfy the Maurer--Cartan structure equation $2d\omega^{I}  -  \epsilon^{I}\,_{JK}   \omega^{J} \wedge \omega^{K} =0$.
At f\/irst order, for a~generic perturbation, let us def\/ine
\begin{gather*}
\tilde\omega_f^I = \f12\int_f\epsilon^I\,_{JK} \tilde\omega^J\wedge\tilde\omega^K=\omega_f^I+ \int_f \epsilon^{I}\,_{JK} \omega^J\wedge\varphi^K =\omega^I_f+\sum_{j\alpha\alpha'} \varphi^{KL}_{j \alpha\alpha'}(t)   \phi_{f KL}^{I j\alpha\alpha'},
\end{gather*}
where
$
\phi_{f KL}^{I j\alpha\alpha'}=
\int_f \epsilon^{I}\,_{JK}  D^j_{\alpha  \alpha'}
 \omega^J\wedge \omega^{L}$.
In particular, if we restrict to the diagonal $j=1$, $\alpha=0$ case,
\begin{gather*}
\tilde\omega_f^I=  \omega_f^I
+  \varphi^{(I)}_{\alpha}(t)  \phi^{(I)}_{f,\alpha},
\qquad \mbox{where}\qquad
\phi^{I}_{f,\alpha}=\int_f \epsilon^{I}\,_{JK}  D^{1}_{0  \alpha}
 \omega^J\wedge \omega^{K}
 \end{gather*}
 are f\/ixed coef\/f\/icients.
Then the relation with the dipole variables can be written as
\begin{gather*}
h_\ell\big(c^I,\varphi^{I}_{\alpha}\big)=\exp\big(c^I\tilde\omega^I_\ell\tau_I\big)\exp\big(\alpha \omega^I_\ell\tau_I\big),
\end{gather*}
which replaces (\ref{varbix}).
Similarly, we can write
\begin{gather*}
\tilde E_\ell^I = \int_\ell (e_I^a+\psi^a_I)\epsilon_{abc}dx^b\wedge dx^c= E^I_\ell
+ 2  \int_\ell \psi_{IJ}   \epsilon^{J}\,_{KL} \omega^K\wedge\psi^L
=E^I_\ell
+  2\sum_{j\alpha\alpha'}  \psi_{IJ}^{j\alpha} \phi_{f, \alpha \alpha'}^J.
\end{gather*}
In particular, if we restrict to the diagonal $j=1$, $\alpha=0$ case,
\begin{gather*}
\tilde E_\ell^I =  E^I_\ell + 2 \psi_{I}^{\alpha}   \phi_{f,\alpha}^I.
\end{gather*}

Notice that now the Gauss constraint do not vanish identically. It can be split into two parts: the homogeneous and the inhomogeneous terms
\begin{gather*}
\mathcal{G}^I=\sum_\ell p^{(I)}\omega_\ell^{(I)}+
2 \psi_{I}^{\alpha}  \sum_\ell \phi^{I}_{f,\alpha}\approx0 .
\end{gather*}
The f\/irst part is the constraint which appears within the Bianchi IX framework and vanishes identically  because of the Stokes theorem. The second gives
three conditions on the inhomogeneous perturbations to the electric f\/ields $\tilde E_\ell$.

\section[The $\U(N)$ framework]{The $\boldsymbol{\U(N)}$ framework}\label{section3}

A number of technical problems are still open in the canonical framework for LQG.
Among these is to determine a form of the dynamics with a fully satisfactory
semiclassical limit (but see the recent developments in~\cite{Domagala:2010nx}), and the dif\/f\/iculty of identifying a homogeneous
sector yielding the Loop Quantum Cosmology (LQC) formalism. A novel point of view
on these problems has been developed in a number of recent papers
\cite{Freidel:2009ck,Freidel:2010tt, Girelli:2005ii}. This is based on the
identif\/ication of a $\U(N)$ symmetry in the Hilbert space of LQG intertwiners with
$N$ legs and f\/ixed total area. In this Section we illustrate the basis of this
approach, called the $\U(N)$ framework, and we introduce the simplest
nontrivial system where  this approach prove useful.  This system is again
based on a 2-node graph
as in the previous Section, but now with an arbitrary number $L$ of links. The
relevant symmetry will be given by the group $\U(N)$, where $N=L$
 \cite{Borja:2010gn,Borja:2010hn,Borja:2010rc}. This framework allows
us to introduce a specif\/ic dynamics for the 2-node graph and a convincing
def\/inition of homogeneity and isotropy. In this sense, the $\U(N)$ framework
may prove a useful tool for addressing the problems mentioned above.

The $\U(N)$ framework has been developed in various directions. Recently,
it was given a nice interpretation in terms of
spinors, where intertwiner spaces can be  reinterpreted as
the product of the quantization of spinors model~\cite{Livine:2011gp}.
Other directions of research are the def\/inition of coherent
states~\cite{Freidel:2010tt}, the reinterpretation of this framework
in terms of holomorphic functions~\mbox{\cite{Borja:2010rc,Livine:2011gp}},
and the study of the simplicity constraints which appear in the spinfoam
models~\cite{Dupuis:2010iq}.  In this review we focus on
the basis of the $\U(N)$ framework, the complete treatment
of the 2-node model and the possibility of understanding it as a
classical system written in terms of spinors.

\subsection[Introduction to the $\U(N)$ framework]{Introduction to the $\boldsymbol{\U(N)}$ framework}

The $\U(N)$ symmetry def\/ines a  framework to  investigate
the structure of the Hilbert space of the intertwiners, which are the
building blocks of spinnetwork states.

We consider f\/irst the space of $N$-valent intertwiners: the
space of $\SU(2)$ invariant tensors of~$N$ spins ($\SU(2)$
representations). Such intertwiners can be thought dually as a
region of 3d space with a (topologically) spherical boundary
punctured by the~$N$ legs of the intertwiners. The boundary
surface is made of~$N$ elementary patches, whose areas are determined
by the spins carried by the intertwiner legs.

The Hilbert space of the intertwiner space for the group $\SU(2)$ is
def\/ined as
\begin{gather*}
\cH_{j_1,\dots,j_N} \equiv  \textrm{Inv}[V^{j_1}\otimes\cdots \otimes
V^{j_N}] ,
\end{gather*}
where $V^{j_i}$ are the irreducible representation spaces associated
to the spin  $j_1,\dots,j_N$.

The key idea of the $\U(N)$ formalism is to consider the space formed
by all the  intertwiners with~$N$ legs and f\/ixed
total sum of the spin numbers $J=\sum_i j_i$  (related with
the total area of the boundary surface). That is
\begin{gather*}
\cH_N^{(J)} \equiv \bigoplus_{\sum_i
j_i=J}\cH_{j_1,\dots,j_N}.
\end{gather*}

{\sloppy It can be shown that the intertwiner space $\cH_N^{(J)}$ carries an
irreducible representation of~$\U(N)$~\cite{Freidel:2009ck}; and
the full space $\cH_N \equiv \bigoplus_{J }\cH^{(J)}$ can be endowed with a Fock space structure
with creation and annihilation operators compatible with the $\U(N)$
action~\cite{Freidel:2010tt}. In the following, we review the basics
of this construction.

}

The f\/irst step to arrive to the $\U(N)$ framework is to make use of
the well known Schwinger representation of the $\su(2)$ algebra.
This representation consists on describing the generators of
$\su(2)$ in terms of a pair of uncoupled harmonic oscillators. We
introduce $2N$ oscillators with operators~$a_i$,~$b_i$ with~$i$ running from $1$ to $N$ (a pair of uncoupled harmonic
oscillators for each leg of the intertwiner), satisfying
\begin{gather*}
[a_i,a^\dag_j]=[b_i,b^\dag_j]=\delta_{ij} ,\qquad
[a_i,b_j]=0.
\end{gather*}
The local $\su(2)$ generators acting on each
leg $i$ are def\/ined as quadratic operators
\begin{gather*}
J^z_i=\f12(a^\dag_i a_i-b^\dag_ib_i),\qquad J^+_i=a^\dag_i
b_i,\qquad J^-_i=a_i b^\dag_i,\qquad E_i=(a^\dag_i a_i+b^\dag_ib_i).
\end{gather*}
The $J_i$'s constructed in this way satisfy the
standard commutation relations of the $\su(2)$ algebra while the
total energy~$E_i$ is a Casimir operator
\begin{gather*}
[J^z_i,J^\pm_i]=\pm J^\pm_i,\qquad [J^+_i,J^-_i]=2J^z_i,\qquad
[E_i,\vec{J}_i]=0.
\end{gather*}
The operator $E_i$ is the total energy carried by the pair of
oscillators~$a_i$, $b_i$ and gives~$2j_i$, namely twice the spin,  of the
corresponding~$\SU(2)$-representation. Indeed, we can easily express
the standard~$\SU(2)$ Casimir operator in terms of this energy
\begin{gather*}
\vJ_i^2 =  \f{E_i}2\left(\f{E_i}2+1\right) =
\f{E_i}4\left({E_i}+2\right).
\end{gather*}
In LQG the spin $j_i$ is related to
the area associated to the leg $i$ of the intertwiner. Notice that in this
context the most natural ordering of the area operator is the one
given by the Casimir~$E_i/2$.

Our goal is to construct operators acting on the
Hilbert space of intertwiners. In other words, we look for operators
invariant under global~$\SU(2)$ transformations generated by
$\vJ \equiv \sum_i \vJ_i$. The key result, which is the starting
point of the $\U(N)$ formalism, is that we can identify quadratic
invariant operators acting on pairs of (possibly equal) legs $i$, $j$
\cite{Freidel:2009ck, Girelli:2005ii}
\begin{gather*}
 E_{ij}=a^\dag_ia_j+b^\dag_ib_j, \qquad E_{ij}^\dag=E_{ji},\qquad
 F_{ij}=(a_i b_j - a_j b_i),\qquad F_{ji}=-F_{ij}.
\end{gather*}
These operators $E$, $F$, $F^\dag$ form a closed algebra:
\begin{gather}
{[}E_{ij},E_{kl}] =
\delta_{jk}E_{il}-\delta_{il}E_{kj},\nonumber\\
{[}E_{ij},F_{kl}]  =  \delta_{il}F_{jk}-\delta_{ik}F_{jl},\qquad
{[}E_{ij},F_{kl}^{\dagger}] =
\delta_{jk}F_{il}^{\dagger}-\delta_{jl}F_{ik}^{\dagger}, \nonumber\\
{[}F_{ij},F^{\dagger}_{kl}]  =  \delta_{ik}E_{lj}-\delta_{il}E_{kj}
-\delta_{jk}E_{li}+\delta_{jl}E_{ki}
+2(\delta_{ik}\delta_{jl}-\delta_{il}\delta_{jk}), \nonumber\\
{[}F_{ij},F_{kl}]  =  0,\qquad {[}
F_{ij}^{\dagger},F_{kl}^{\dagger}] =0.\label{commEF}
\end{gather}
The commutators of the $E_{ij}$ operators form a $\u(N)$-algebra
(hence the name of the $\U(N)$ framework). The
diagonal operators are equal to the
energy on each leg, $E_{ii}=E_i$. The value of the total energy
$E \equiv \sum_i E_i$ gives twice the sum of all spins
$2\times\sum_i j_i$, i.e.\ twice the total area.

The $E_{ij}$-operators change the energy/area carried by each leg,
while still conserving the total energy, while the operators
$F_{ij}$ (resp.\ $F^\dag_{ij}$)  decrease (resp.\ increase) the
total area $E$ by 2
\begin{gather*}
[E,E_{ij}]=0,\qquad [E,F_{ij}]=-2F_{ij},\qquad
[E,F^\dag_{ij}]=+2F^\dag_{ij}.
\end{gather*}
This suggests to decompose the Hilbert space of $N$-valent
intertwiners into subspaces of f\/ixed area
\begin{gather*}
\cH_N=\bigoplus_{\{j_i\}} \inv\left[\otimes_{i=1}^NV^{j_i}\right]
=\bigoplus_{J\in\N}\bigoplus_{\sum_ij_i=J}
\inv\left[\otimes_{i=1}^NV^{j_i}\right] =\bigoplus_J
\cH_N^{(J)},
\end{gather*}
where $V^{j_i}$ denote the Hilbert space of the irreducible
$\SU(2)$-representation of spin $j_i$, spanned by the states  of the
oscillators $a_i$, $b_i$ with f\/ixed total energy $E_i=2j_i$.

It was proven in \cite{Freidel:2009ck} that each subspace $\cHNJ$ of
$N$-valent intertwiners with f\/ixed total area~$J$ carries an
irreducible representation of $\U(N)$ generated by the $E_{ij}$
operators. The opera\-tors~$E_{ij}$ allow to navigate from state
to state within each subspace $\cHNJ$. On the other hand, the
operators~$F_{ij}$, $F^\dag_{ij}$ allow to go from one subspace
$\cHNJ$ to the next $\cHN^{(J\pm 1)}$, thus endowing the full space
of $N$-valent intertwiners with a Fock space structure with creation
operators $F^\dag_{ij}$ and annihilation operators $F_{ij}$.

Finally, the whole set of operators
$E_{ij}$, $F_{ij}$, $F^\dag_{ij}$ satisfy quadratic constraints
\cite{Borja:2010gn}, $\forall\, i,j$:
\begin{gather}
 \sum_k E_{ik}E_{kj}=E_{ij} \left(\f
E2+N-2\right),\label{constraint0}\\
 \sum_k F^\dagger_{ik}E_{jk} =
F^\dagger_{ij}  \frac{E}{2}, \qquad
\sum_k E_{jk} F^\dagger_{ik} = F^\dagger_{ij}\left(\frac{E}{2}+N-1\right),\label{constraint1}\\
 \sum_k E_{kj}F_{ik} = F_{ij}  \left(\frac{E}{2}-1\right), \qquad
\sum_k F_{ik} E_{kj}  = F_{ij}\left(\frac{E}{2}+N-2\right),\label{constraint2}\\
 \sum_k F^\dagger_{ik}F_{kj} = E_{ij}
\left(\frac{E}{2}+1\right),\qquad \sum_k F_{kj}F^\dag_{ik} =
(E_{ij}+2\delta_{ij})
\left(\frac{E}{2}+N-1\right) .\label{constraint3}
\end{gather}
As already noticed in \cite{Borja:2010gn} and extended in~\cite{Borja:2010rc}, these relations look a lot like constraints on
the multiplication of two matrices~$E_{ij}$ and~$F_{ij}$. We will
explore this fact at the end of this section, but f\/irst we
discuss the 2-node model from the~$\U(N)$ perspective.

\subsection{Hilbert space for the 2-node model}

\begin{figure}[h]
\centering

\includegraphics[height=40mm]{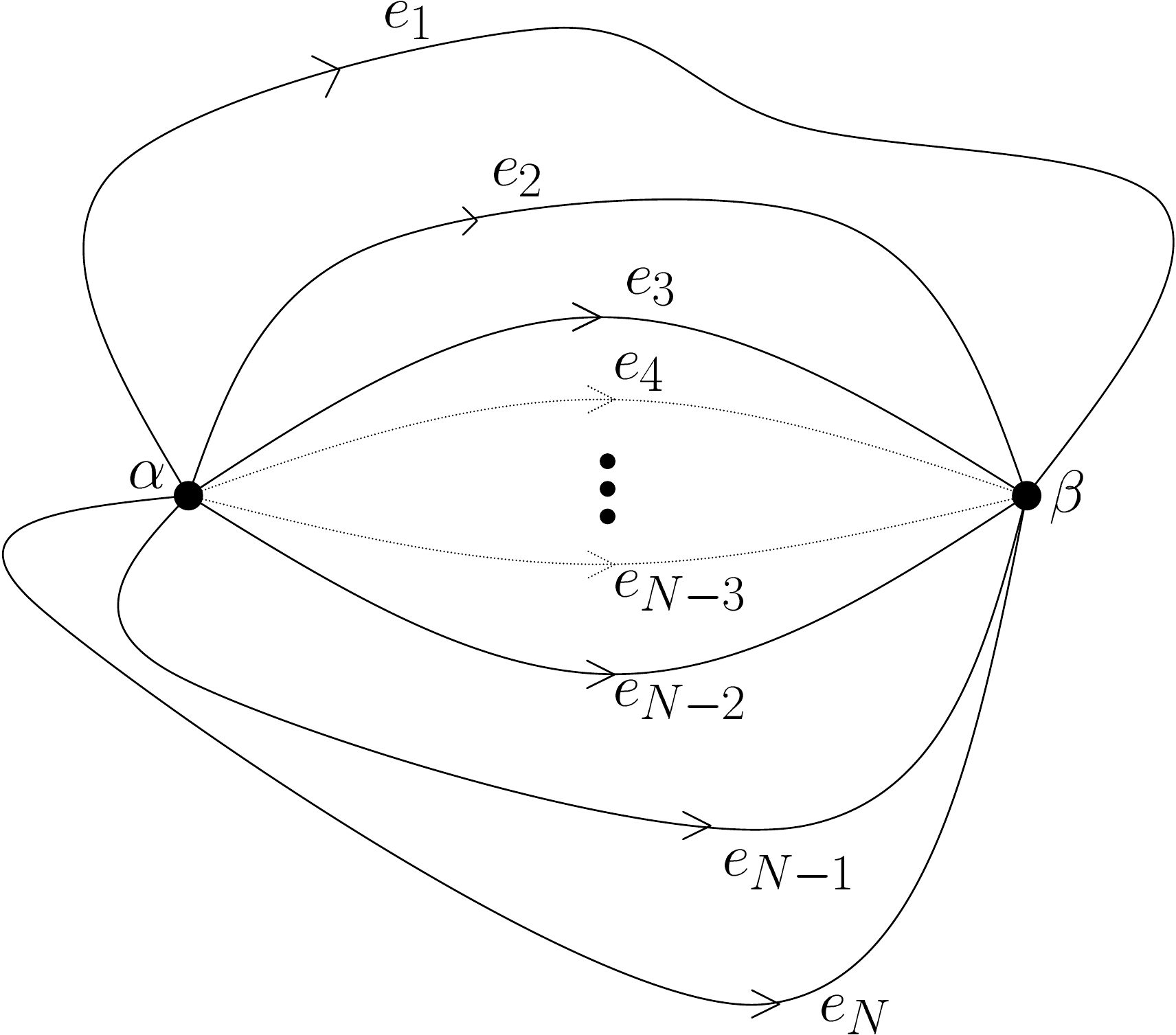}
\caption{The 2-node graph: the two nodes $\alpha$ and $\beta$ are
linked by $N=L$ links. \label{2vertex}}
\end{figure}

We are going to establish the main steps to construct the Hilbert
space of the 2-node model. In this case, we need to deal with two
intertwiner spaces of $N$ legs each, as shown in Fig.~\ref{2vertex}.

Naming the two nodes $\alpha$ and $\beta$, we have two intertwiner
spaces attached respectively to $\alpha$ and $\beta$ with
operators $\Ea_{ij}$, $\Fa_{ij}$ and $\Eb_{ij}$, $\Fb_{ij}$. The
total Hilbert space of these uncoupled intertwiners is the tensor
product of the two spaces of $N$-valent intertwiners
\begin{gather*}
\cH_{\otimes 2}  =  \cH_N\otimes\cH_N
=\bigoplus_{J_\alpha,J_\beta}\cH_N^{(J_\alpha)}\otimes\cH_N^{(J_\beta)}
=\bigoplus_{\{j_i^\alpha,j_i^\beta\}}
\cH_{j_1^\alpha,\dots,j_N^\alpha}\otimes\cH_{j_1^\beta,\dots,j_N^\beta}.
\end{gather*}
This space carries two decoupled $\U(N)$-actions, one acting on the
intertwiner space associated to the node~$\alpha$ and the other
acting on~$\beta$. However, the two intertwiner spaces are not
independent. There are matching conditions
unique $\SU(2)$ representation, thus the spin on that link must be
the same seen from~$\alpha$ or~$\beta$, i.e.\ $j_i^\alpha=j_i^\beta$.
This translates into the fact that the oscillator energy for~$\alpha$ on the leg~$i$ must be equal to the energy for~$\beta$ on
its $i$-th leg
\begin{gather*}
\cE_i \equiv \Ea_i -\Eb_i  = 0.
\end{gather*}
Obviously, this set of conditions is stronger than requiring that
the total area of~$\alpha$ is the same as~$\beta$, even though this
is a necessary condition. Then, the Hilbert space of spinnetwork
states on the 2-node graph is much smaller than the decoupled
Hilbert space $\cH_{\otimes 2}$
\begin{gather*}
\dcH\equiv\bigoplus_{\{j_i\}}\cH_{j_1,\dots,j_N}^{(\alpha)}\otimes
\cH_{j_1,\dots,j_N}^{(\beta)}.
\end{gather*}
In order to def\/ine consistent operators acting on $\dcH$, we have to
check that they commute (at least weakly) with the matching
conditions~$\cE_i$, in addition to the condition that they need to
be invariant under global~$\SU(2)$ transformations. It is possible
to construct such operators deforming consistently the boundary
between~$\alpha$ and~$\beta$. We introduce the following operators,
mixing actions on the two nodes as required
\begin{gather}
\label{operatorsef}
e_{ij}\equiv\Ea_{ij}\Eb_{ij},\qquad
f_{ij}\equiv\Fa_{ij}\Fb_{ij},\qquad
f^\dag_{ij}\equiv\Fa_{ij}{}^\dag\Fb_{ij}{}^\dag ,
\end{gather}
and we check that they commute with the matching conditions
\begin{gather*}
\forall\, i,j,k,\qquad [\cE_k,e_{ij}]  = [\cE_k,f_{ij}]  =  0.
\end{gather*}
Calling $E=\sum_i \Ea_i=\sum_i \Eb_i$ the operator giving (twice)
the total boundary area on our Hilbert space $\dcH$ satisfying the
matching conditions, the operators $e_{ij}$'s preserve the boundary
area while the $f_{ij}$'s will (as it is expected by construction)
modify it
\begin{gather*}
[E,e_{ij}]=0,\qquad [E,f_{ij}]=-2f_{ij},\qquad
[E,f^\dag_{ij}]=+2f^\dag_{ij}.
\end{gather*}
More precisely, the operator $e_{ij}$ increases the spin on the
$i$-th link by $+\f12$ and decreases the spin of the $j$-th link.
The operator $f_{ij}$ decreases both spins on the links~$i$ and~$j$,
while its adjoint~$f^\dag_{ij}$ increases both spins by~$\f12$.
These operators generate the deformations of the boundary surface,
consistently with both the~$\SU(2)$ gauge invariance and the
matching conditions imposed by the graph combinatorial structure.
They are natural building blocks for the dynamics of spin
network states on the 2-node graph.

Nevertheless, the operators $e_{ij}$, $f_{ij}$ and $f^\dag_{ij}$
are not enough to generate the whole Hilbert space~$\dcH$ of spin
network states from the vacuum state~$|0\ra$. Indeed, they are
symmetric in $\alpha\leftrightarrow \beta$ and will only generate
states symmetric under the exchange of the two nodes. In order to
generate the whole space of gauge invariant operators on the 2-node
graph, we also need operators that act on a single node and change
states without af\/fecting neither the representations on the edges,
nor the intertwiner state living at the other node. Natural
candidates for such operators acting on
$\cH_{j_1,\dots,j_N}^{(\alpha)}$ are the $\Ea_{ij}\Ea_{ji}$. They
change the intertwiner living at the node $\alpha$ without changing
the intertwiner living at $\beta$. Combining such local operators
$\Ea_i$, $\Ea_{ij}\Ea_{ji}$ and $\Eb_i$, $\Eb_{ij}\Eb_{ji}$ with the
coupled symmetric operators $e_{ij}$, $f_{ij}$, $f^\dag_{ij}$ allow to go
between any two states in the Hilbert space $\dcH$ and thus generate
all gauge invariant operators on the 2-node graph.

\subsection{Symmetry reduction and cosmological analogies}
\label{uns}

We now consider a symmetry reduction in our simple model that
yields a homogeneous and isotropic sector. More specif\/ically, we look for states in the Hilbert
space $\dcH$ invariant under a ``global'' $\U(N)$ symmetry
generated by a generalization of the matching conditions $\cE_k$,
and which takes into account operators~$E_{ij}$ acting on both nodes
of the graph. Let us explain here the construction of the generators
of this new $\U(N)$ symmetry and its action.

The matching conditions $\cE_k$ break the $\U(N)$-actions on both
nodes $\alpha$ and $\beta$. Nevertheless, we can see that the
$\cE_k$ generate a $\U(1)^N$ symmetry and that they are part of a
larger $\U(N)$ symmetry algebra. We introduce the operators
\begin{gather*}
\cE_{ij} \equiv  \Ea_{ij}-\Eb_{ji}  = \Ea_{ij}-\big(\Eb_{ij}\big)^\dag.
\end{gather*}
It is straightforward to compute their commutation relations and
check that these operators form a ~$\u(N)$ algebra
\begin{gather*}
[\cE_{ij},\cE_{kl}] = \delta_{jk}\cE_{il}-\delta_{il}\cE_{kj}.
\end{gather*}
The diagonal operators are exactly the matching conditions
$\cE_{kk}=\cE_k$ and generate the Cartan Abelian subalgebra of
$\u(N)$.

These $\cE_{ij}$'s generate $\U(N)$ transformations on the two
intertwiner system. By construction, they act in~$\cH_{\otimes 2}$
as $(U,\bar{U})$ with the transformation on~$\beta$ being the
complex conjugate of the transformation on~$\alpha$.

Two remarks are in order when computing the commutator between the
matching condition and these new~$\u(N)$ generators,
\begin{gather*}
[\cE_{ij},\cE_k]  =  \delta_{jk}\cE_{ik}-\delta_{ik}\cE_{kj}  =
(\delta_{jk}-\delta_{ik})\cE_{ij} .
\end{gather*}
On the one hand, we notice that the operators~$\cE_{ij}$ are not
fully compatible with the matching conditions and they do not act on
the 2-node Hilbert space~$\dcH$. Thus they do not generate a~nontrivial $\U(N)$-action on $\dcH$. On the other hand, we can look
for vectors in  $\dcH$ which  are invariant under this~$\U(N)$
action, $\cE_{ij}\,|\psi\ra=0$ for all~$i$,~$j$. In particular, they
will satisfy the matching conditions (given by the special case
$i=j$).

Following this line of thought, we introduce the subspace of spin
network states which are invariant under this $\U(N)$-action
\begin{gather*}
\cHi \equiv  {\rm Inv}_{\U(N)}\left[\dcH\right]  =
{\rm Inv}_{\U(N)}\left[\cH_{\otimes 2}\right]  =
{\rm Inv}_{\U(N)}\left[\bigoplus_{J_\alpha,J_\beta}\cH^{(J_\alpha)}_N\otimes\cH^{(J_\beta)}_N\right].
\end{gather*}

Now taking into account that the spaces $\cH^{(J)}_N$ are
irreducible $\U(N)$-representations~\cite{Freidel:2009ck}, requiring
$\U(N)$-invariance imposes that the two representations for the two
nodes are the same, $J_\alpha=J_\beta$, but furthermore there exists
a unique invariant vector in the tensor product
$\cH^{(J)}_N\otimes\cH^{(J)}_N$. We will call this unique invariant
vector $|J\ra$ and we will construct it explicitly in terms of the
operators $e_{ij}$ and $f_{ij}$ in the next section.

It is important to notice that imposing $\U(N)$-invariance on our
2-node system, we obtain a~single state $|J\ra$ for every total
boundary area $J$
\begin{gather*}
\cHi = \bigoplus_{J\in\N} \C\,|J\ra.
\end{gather*}
We def\/ine these $|J\ra$ states as the homogeneous and isotropic
states of the model. The physical motivation behind this def\/inition
is that $\U(N)$ invariance is restricting our system to states which
are not sensitive to area-preserving deformations of the boundary
between~$\alpha$ and~$\beta$. They are isotropic in the sense that
all directions (i.e.\ all links) are equivalent and the state only
depends on the total boundary area, and they are homogeneous in the
sense that the quantum state is the same at every point of space,
i.e.\ at both nodes $\alpha$ and $\beta$ of our 2-node graph.

This allows to realize the reduction at the quantum level to the
isotropic/homogeneous subspace by a straightforward $\U(N)$-group
averaging. This opens the possibility of applying this logic to Loop
Quantum Cosmology, which is based on a symmetry reduction at the
classical level  and a quantization {\it \`a la} loop of this
reduced phase space. As it will be explained in the next section,
the dynamics that we propose for the $\U(N)$ invariant sector has
also strong analogies with the evolution operator used in LQC.

\subsection{Dynamics for the 2-node model}

In this section, we will def\/ine a consistent dynamics based on the
$\U(N)$ invariance, restricting ourselves to the subspace of
homogeneous/isotropic states described previously. In particular,
such dynamics will automatically be consistent with the matching
conditions.

\subsubsection[The algebra of $\U(N)$ invariant operators]{The algebra of $\boldsymbol{\U(N)}$ invariant operators}

Before proposing a Hamiltonian operator for this system, we explore
here the dif\/ferent $\U(N)$-invariant operators that we can
construct. The most obvious one is the total boundary area operator
$E$ itself. It is def\/ined as $E=\Ea=\Eb$ on the space $\dcH$ of spin
network states satisfying the matching condition. It is direct to
check that it commutes with the $\u(N)$ generators
\begin{gather*}
[\cE_{ij},\Ea]=[\Ea_{ij},\Ea]  =  0  =  -[\Eb_{ji},\Eb]
=[\cE_{ij},\Eb].
\end{gather*}
This total area operator is clearly diagonal in the basis $|J\ra$,
\[
E\,|J\ra  =  2J\,|J\ra.
\]
Now, we need operators that could create dynamics on the space
$\cHi$ by inducing transitions between states with dif\/ferent areas.
To this purpose, we use the operators $e_{ij}$, $f_{ij}$ (equation~\eqref{operatorsef}) and introduce the unique linear combinations that
are $\U(N)$-invariant
\begin{gather*}
e\equiv\sum_{ij}e_{ij}=\sum_{ij}\Ea_{ij}\Eb_{ij},\qquad
f\equiv\sum_{ij}f_{ij}=\sum_{ij}\Fa_{ij}\Fb_{ij}.
\end{gather*}
They obviously commute with the matching conditions since each
operator~$e_{ij}$ and~$f_{ij}$ does. An important feature of these
new operators, is that they are quadratic in~$E$ and/or~$F$
separately, so in this case their action is the same in both nodes.
This actually ensure that the matching conditions hold.

It is convenient, for computational purposes, to introduce a shifted
operator $\tl{e}\equiv e+2(E+N-1)$. Then, using the quadratic
constraints (\ref{constraint1})--(\ref{constraint3}) satisf\/ied by the
operators $E_{ij}$ and $F_{ij}$, we can show that~$\tl{e}$,~$f$ and~$f^\dag$ form a simple algebra
\begin{gather*}
 [\te,f ] = -2(E+N+1)f ,\qquad
 [\te,f^{\dagger} ] = 2f^{\dagger}(E+N+1),\qquad
 [f,f^{\dagger} ] = 4(E+N)\te .
\end{gather*}
Written as such, it resembles to a $\sl_2$ Lie algebra up to the
factors in $E$, which is an operator and not a
constant\footnote{This is very similar to the $\sl_2$ algebra
$\Ea_\rho$, $\Fa_{\bz}$, $\Fa_{\bz}{}^\dag$ def\/ined in~\cite{Freidel:2010tt} and used to build the~$\U(N)$ coherent
states.}.

We can use $f^\dag$ as a creation operator. Thus we introduce the
states
\begin{gather*}
|J\ra_{\rm un}\equiv f^{\dagger J}|0\ra  =
\left(\sum_{ij}\Fa_{ij}{}^\dag\Fb_{ij}{}^\dag\right)^J\,|0\ra,
\end{gather*}
where the index ${\rm un}$ stands for unnormalized. Since both the
operator $f^\dag$ and the vacuum state~$|0\ra$ are
$\U(N)$-invariant, it is clear that the states $|J\ra_{\rm un}$ are also
invariant under the $\U(N)$-action. Moreover, it is easy to check
that they are eigenvectors of the total area operator:
\[
E\,|J\ra_{\rm un}=2J\,|J\ra_{\rm un},
\]
so that they provide a basis for our Hilbert space $\cHi$ of
homogeneous states. It is possible also to work with normalized
states def\/ined as
\begin{gather*}
|J\ra \equiv  \f{1}{2^J J!(J+1)! \sqrt{D_{N,J}}}\,|J\ra_{\rm un},
\end{gather*}
in terms of the dimension $D_{N,J}$ of the intertwiner space
$\cH_N^{(J)}$~\cite{Freidel:2009ck} (see~\cite{Borja:2010gn} for
details).

The action of all $\tl{e}$, $f$, $f^\dag$ operators over these normalized
states is always quadratic in~$J$:
\begin{gather*}
\te\,|J\ra  =  2(J+1)(N+J-1)|J\ra ,\\
 f\,|J\ra  = 2\sqrt{J(J+1)(N+J-1)(N+J-2)}\,|J-1\ra,\\
  f^\dag\,|J\ra = 2\sqrt{(J+1)(J+2)(N+J)(N+J-1)}\,|J+1\ra.
\end{gather*}

From here, we see that it is possible to introduce renormalized
operators that truly form a~$\sl_2$ algebra. We def\/ine the new
operators:
\begin{gather*}
Z \equiv  \f{1}{\sqrt{E+2(N-1)}}  \te  \f{1}{\sqrt{E+2(N-1)}} ,\\
X_- \equiv  \f{1}{\sqrt{E+2(N-1)}}  f   \f{1}{\sqrt{E+2(N-1)}} ,\\
X_+ \equiv
\f{1}{\sqrt{E+2(N-1)}}  f^\dag  \f{1}{\sqrt{E+2(N-1)}} .
\end{gather*}
Notice that the inverse square-root is well-def\/ined since $E+2(N-1)$
is Hermitian and strictly positive as soon as~$N\ge 2$. These
operators are still $\U(N)$-invariant since $E$ is invariant too and
we also  have the Hermiticity relations, $Z^\dag=Z$ and
$X_-^\dag=X_+$. Moreover, it is direct to compute the action of
these renormalized operators on our $|J\ra$ basis states.

To conclude, we can extract two important points
\begin{itemize}\itemsep=0pt
\item The algebraic structure of the $\U(N)$-invariant space $\cHi$
of homogeneous states forms an irreducible unitary representation of
$\sl(2,\R)$. The basis vectors $|J\ra$ can be obtained by iterating
the action of the creation/raising operator $f^\dag$  (or $X_+$) on
the vacuum state $|0\ra$.

\item This algebraic structure does not depend at all on the number of links $N$.
Therefore, while working on homogeneous states, $N$ might have a
physical meaning but it is not a relevant parameter mathematically.
On the other hand, we expect it to become highly relevant when
leaving the $\U(N)$-invariant subspace and studying inhomogeneities.
\end{itemize}

\subsubsection{The Hamiltonian}

In order to study the dynamics on this 2-node graph we propose the
simplest $\U(N)$-invariant ansatz for a Hamiltonian operator
\begin{gather*}
H  \equiv  \eta\te+ (\sigma f +\bar{\sigma} f^\dag).
\end{gather*}
As explained above, the operator $\te$ does not af\/fect the total
boundary area, $[E,\te]=0$, while the operators $f$ and $f^\dag$
respectively shrink and increase this area, $[E,f]=-2f$ and
$[E,f^\dag]=+2f^\dag$. The coupling~$\eta$ is real while $\sigma$
can be complex a priori, so that the operator $H$ is Hermitian. We
can relate this Hamiltonian operation to the action of holonomy
operators acting on all the loops of the 2-node graph~\cite{Borja:2010gn}. From this point of view, our proposal is very
similar to the standard ansatz for the quantum Hamiltonian
constraint in LQG~\cite{Thiemann:1996aw} and LQC (e.g.~\cite{Ashtekar:2006es, Kaminski:2008td}).

This Hamiltonian is quadratic in $J$ and we can give its explicit
action on the basis states of~$\cHi$:
\begin{gather}
H|J\ra  =  \sigma \,2\sqrt{J(J+1)(N+J-1)(N+J-2)}\,|J-1\ra
 +\eta 2(J+1)(N+J-1)\,|J\ra\nonumber\\
\phantom{H|J\ra  =}{}+\bar{\sigma} 2\sqrt{(J+1)(J+2)(N+J)(N+J-1)}\,|J+1\ra ,\label{83}
\end{gather}
the details of the spectral properties of this Hamiltonian can be
found in \cite{Borja:2010gn}. We observe in the spectral analysis
that it presents three dynamical regimes depending on the value of
the couplings that can be put in the same footing as the three
regimes in LQC given by the sign of the cosmological constant~\cite{Borja:2010gn}.
Moreover, we can see that the structure of the Hamiltonian~\eqref{83} is analogous to the one given in equation \eqref{din2}, with a reparametrization by a factor~2 of the label $J$. Notice that both labels $\mu$ and $J$ are area values, the coef\/f\/icients of both equations are quadratic and both of them have contributions of three states labelled by $\mu$, $\mu-2$ and $\mu+2$ in the case of equation~\eqref{din2} and $J$, $J-1$, $J+1$ in the case of \eqref{83}.

The ansatz given above is the most general $\U(N)$-invariant
Hamiltonian (allowing only elementary changes in the total area), up
to a renormalization by a $E$-dependent factor. Therefore, we can
also propose renormalized Hamiltonian operators based on the
renormalized operators considered in the previous section. For
instance, we can def\/ine
\begin{gather*}
\hh \equiv  \f{1}{\sqrt{E+2(N-1)}} H \f{1}{\sqrt{E+2(N-1)}}
 =  \eta Z+(\sigma X_-+\bar{\sigma} X_+) \in \sl_2 .
\end{gather*}
We remark the fact that we study the action of these Hamiltonian
operators, $H$ and~$\hh$, on the $\U(N)$-invariant space~$\cHi$;
nevertheless they are generally well-def\/ined on the whole space of
spinnetwork states~$\dcH$. Also, that the dynamics on the
homogeneous sector does not depend mathematically on the parameter
$N$ giving the number of links of the graph, as it does not appear
in the action of the renormalized Hamiltonian~$\hh$.

The main characteristic of $\hh$ is that it is an element in the Lie
algebra $\sl_2$ and its coef\/f\/icients are linear in the variable $J$
at leading order. Thus we know its spectral properties from the
representation theory of~$\sl_2$. The important point to underline
here is that the three coupling regimes for the renormalized
Hamiltonian~$\hh$ are exactly the same as for the original
Hamiltonian~$H$.  This is very similar to the interplay between the
{\it  evolution operator} $\hat{\Theta}$ and the gravitational
contribution to the Hamiltonian constraint $\hat{C}_{\textrm{grav}}$
(see e.g.~\cite{Kaminski:2008td}) in LQC. At the end of this
section, we will obtain the classical counterpart of these
Hamiltonians and we will be able to solve the equations of motion
for~$\hh$ exactly.

\subsection{Classical setting: formulation of the 2-node model in terms of spinors}

One of the most interesting aspects of the $\U(N)$-framework is
that it can be rewritten in terms of spinors rather straightforward~\cite{Borja:2010rc,Livine:2011gp}. This fact is natural in this
setting because the operators in the $\U(N)$-formalism can be seen as
the quantization of a classical spinorial model. This relationship
may lead us to a better understanding of the geometrical meaning of
the spinnetwork states in LQG and can also help us in the quest of
a well def\/ined semi-classical limit for the full theory.

Another remarkable point is the direct relation with the work by L.~Freidel and S.~Speziale so called ``twisted geometries''~\cite{Freidel:2010uq,Freidel:2010bw}. Within this point of view, it
was shown that the classical phase space of loop gravity on a given
graph can be understood as a classical spinor model unravelling the
connection between spinnetworks and the discrete geometry, mainly
the Regge theory.  This could be a key ingredient in order to shed
light about the physical meaning of the spinfoam approach which
treats the dynamics of the spinnetworks.

In this section we present the main elements allowing the
recast of the $\U(N)$-framework in terms of spinors, showing how this
is related with the usual $\SU(2)$ intertwiners in LQG.  First of
all, we def\/ine the classical spinor phase space.  Later on, we
propose a classical theory based on an action principle which
actually gives us that phase space.  After this classical step we
perform the quantization, choosing a specif\/ic polarization based on
certain spinorial holomorphic functionals and we will f\/ind that we
obtain the correct intertwiner Hilbert space. Finally, we will
discuss some interesting topics about the dynamics in our 2-node
model using spinors and some points of contact with LQC.

\subsubsection{Spinors and notation}

In this part, we introduce the spinors and the related useful
notations that we will be using in the rest of the section
\cite{Borja:2010rc,Dupuis:2010iq,Freidel:2010tt,Freidel:2010bw,Livine:2011gp}.
Given a spinor $z$
\[
|z\ra=\begin{pmatrix} z^0\\ z^1\end{pmatrix}, \qquad \la z|=\begin{pmatrix} \bar{z}^0 &
 \bar{z}^1\end{pmatrix},
\]
it is well known that there is a geometrical 3-vector $\vec{V}(z)$,
def\/ined from the projection of the $2\times 2$ matrix $|z\ra\la z|$
onto Pauli matrices $\sigma_a$ (taken Hermitian and normalized so
that $(\sigma_a)^2=\id$)
\begin{gather*} 
|z\ra \la z| = \f12 \big( {\la z|z\ra}\id  +
\vec{V}(z)\cdot\vec{\sigma}\big).
\end{gather*}
It is straightforward to compute the norm and the components of this
vector in terms of the spinors
\begin{gather*}
|\vec{V}(z)| = \la z|z\ra= \big|z^0\big|^2+\big|z^1\big|^2,\\
V^z=\big|z^0\big|^2-\big|z^1\big|^2,\qquad V^x=2\,{\rm Re} \big(\bar{z}^0z^1\big),\qquad
V^y=2\,{\rm Im} \big(\bar{z}^0z^1\big).
\end{gather*}
Also, it is important to notice that the spinor $z$ is entirely
determined by the corresponding 3-vector $\vec{V}(z)$ up to a global
phase. We can give the reverse map
\begin{gather*}
z^0=e^{i\phi} \sqrt{\f{|\vec{V}|+V^z}{2}},\qquad
z^1=e^{i(\phi-\theta)} \sqrt{\f{|\vec{V}|-V^z}{2}},\qquad
\tan\theta=\f{V^y}{V^x},
\end{gather*}
where $e^{i\phi}$ is an arbitrary phase.

We can also introduce the map duality $\varsigma$ acting on spinors
\begin{gather*}
\varsigma\begin{pmatrix}z^0\\ z^1\end{pmatrix}  =
\begin{pmatrix}-\bar{z}^1\\ \bar{z}^0 \end{pmatrix},
\qquad \varsigma^{2}=-1.
\end{gather*}
This is an anti-unitary map, $\la \varsigma z| \varsigma w\ra= \la
w| z\ra=\overline{\la z| w\ra}$, and we will write the related state
as
\[
|z]\equiv \varsigma  | z\ra,\qquad [z| w] = \overline{\la z|
w\ra}.
\]
This map $\varsigma$ maps the 3-vector $\vec{V}(z)$ onto its
opposite
\begin{gather*}
|z][  z| = \f12 \big({\la z|z\ra}\id -
\vec{V}(z)\cdot\vec{\sigma}\big).
\end{gather*}

Finally considering the setting necessary to describe intertwiners
with $N$ legs, we consider~$N$ spinors~$z_i$ and their corresponding
3-vectors~$\vV(z_i)$. Typically, we can require that the~$N$ spinors
satisfy a closure condition, i.e.\  that the sum of the corresponding
3-vectors vanishes, \mbox{$\sum_i \vec{V}(z_i)=0$}. Coming back to the
def\/inition of the 3-vectors $\vV(z_i)$, the closure condition is
easily translated in terms of $2\times 2$ matrices
\begin{gather*}
\sum_i |z_i\ra \la z_i|=A(z)\id, \qquad\textrm{with}\qquad
A(z)\equiv\f12\sum_i \la z_i|z_i\ra=\f12\sum_i|\vec{V}(z_i)|.
\end{gather*}
This further translates into quadratic constraints on the spinors
\begin{gather*}
\sum_i z^0_i \bar{z}^1_i=0,\qquad \sum_i \left|z^0_i\right|^2=\sum_i
\left|z^1_i\right|^2=A(z). 
\end{gather*}
In simple terms, it means that the two components of the spinors,
$z^0_i$ and $z^1_i$, are orthogonal $N$-vectors of equal norm. In
order to simplify the notation, let us introduce the matrix elements
of the $2\times 2$ matrix $\sum_i |z_i\ra\la z_i|$
\begin{gather*}
\cC_{ab}=\sum_i z^a_i \bz^b_i.
\end{gather*}
Then the unitary or closure conditions are written very simply
\begin{gather*}
\cC_{00}-\cC_{11}=0,\qquad \cC_{01}=\cC_{10}=0.
\end{gather*}

\subsubsection{Phase space and quantization}

We are ready now to describe the phase space in terms of spinors.
This will provide us with the suitable arena to proceed with the
quantization~\cite{Borja:2010rc, Livine:2011gp}.

Then, we f\/irst introduce a simple Poisson bracket on our space of~$N$ spinors
\begin{gather*}
\{z^a_i,\bz^b_j\} \equiv i \delta^{ab}\delta_{ij},
\end{gather*}
with all other brackets vanishing,
$\{z^a_i,z^b_j\}=\{\bz^a_i,\bz^b_j\}=0$. This is exactly the Poisson
bracket for $2N$ decoupled harmonic oscillators.

We expect that the closure conditions generates global $\SU(2)$
transformations on the $N$ spinors. In order to check that, we have
to compute the Poisson brackets between the various components of
the $\cC$-constraints:
\begin{gather}
\{\cC_{00}-\cC_{11},\cC_{01}\}=-2i\cC_{01},\qquad
\{\cC_{00}-\cC_{11},\cC_{10}\}=+2i\cC_{10},\qquad
\{\cC_{10},\cC_{01}\}=i(\cC_{00}-\cC_{11}),\nonumber\\
 \{\tr \cC,\cC_{00}-\cC_{11}\}=\{\tr \cC,\cC_{01}\}=\{\tr
\cC,\cC_{10}\}=0.\label{commC}
\end{gather}
These four components $\cC_{ab}$ do indeed form a closed~$\u(2)$
algebra with the three closure conditions $\cC_{00}-\cC_{11}$,
$\cC_{01}$ and $\cC_{10}$ forming the $\su(2)$ subalgebra. Thus we
will write $\vec{\cC}$ for these three $\su(2)$-generators with
$\cC^z\equiv\cC_{00}-\cC_{11}$ and $\cC^+=\cC_{10}$ and
$\cC^-=\cC_{01}$.
The three closure conditions $\vec{\cC}$ will actually become the
generators $\vJ$ at the quantum level, while the operator $\tr \cC$
will correspond to the total energy/area $E$.

Now, let us def\/ine matrices $M$ and $Q$ in the following way
\begin{gather*}
M_{ij}=\la z_i |z_j \ra=\overline{\la z_j |z_i \ra}, \qquad
Q_{ij}=\la z_j |z_i]=\overline{[ z_i |z_j \ra}=-\overline{[ z_j |z_i
\ra}
\end{gather*}
with
\begin{gather*}
z_i \equiv \begin{pmatrix} \bar{u}_{i1}\sqrt{\lambda}\\
\bar{u}_{i2}\sqrt{\lambda}\end{pmatrix},\qquad \lambda\equiv\tr M/2
\end{gather*}
and $u_{ij}$ elements of a unitary matrix. It is possible to write
this matrices as
\begin{gather*}
 M=\lambda U\Delta U^{-1},\!\qquad  \Delta=\begin{pmatrix} 1 & & \\ &1
& \\ \hline && 0_{N-2} \end{pmatrix} , \!\qquad
 Q=\lambda U\Delta_\eps \tU,\!\qquad \Delta_\eps=\begin{pmatrix} & 1& \\
-1& & \\ \hline & &0_{N-2} \end{pmatrix},
\end{gather*}
where $U$ is a unitary matrix $U^\dag U=\id$.

It is easy to show that indeed (up to a global phase) these matrices
are the most general ones satisfying $M=M^\dag$, ${}^tQ=-Q$ and
the classical analogs to the quadratic constraints satisf\/ied by the
operators $E$ and $F$. On the other hand there is a fundamental
point in this construction which is that the unitarity condition on
the matrices $U$ is equivalent (with the presented def\/inition of the
spinors in terms of the unitary matrix elements) to the closure
conditions on the spinors.

Now, we can also compute the Poisson brackets of the $M_{ij}$ and
$Q_{ij}$ matrix elements:
\begin{gather}
 \{M_{ij},M_{kl}\}=i(\delta_{kj}M_{il}-\delta_{il}M_{kj}),
\qquad \{M_{ij},Q_{kl}\}=i(\delta_{jk}Q_{il}-\delta_{jl}Q_{ik}),\nonumber\\
 \{Q_{ij},Q_{kl}\}=0,\qquad
 \{\bar{Q}_{ij},Q_{kl}\}=i(\delta_{ik}M_{lj}+\delta_{jl}M_{ki}-\delta_{jk}M_{li}-\delta_{il}M_{kj}),\label{commM}
\end{gather}
which reproduces the expected commutators~\eqref{commEF} up to the
$i$-factor. We further check that these variables commute with the
closure constraints generating the $\SU(2)$ transformations
\begin{gather*}
\{\vcC,M_{ij}\}=\{\vcC,Q_{ij}\}=0.
\end{gather*}
Finally, we look at their commutator with $\tr  \cC$
\begin{gather*}
\{\tr \cC,M_{ij}\}=0,\qquad \{\tr \cC,Q_{ij}\}=\bigg\{\sum_k
M_{kk},Q_{ij}\bigg\}= +2i Q_{ij},
\end{gather*}
which conf\/irms that the matrix~$M$ is invariant under the full
$\U(2)$ subgroup and that $\tr \cC$ acts as a dilatation operator
on the~$Q$ variables, or reversely that the $Q_{ij}$ acts as
creation operators for the total energy/area variable $\tr \cC$.

So far, we have been able to characterize the classical phase space
associated to the spinors~$z_i$ and the variables~$M_{ij}$,~$Q_{ij}$.
Then we can now proceed to the quantization. In order to do that, we
introduce the Hilbert spaces~$\cHQJ$ of homogeneous polynomials in
the~$Q_{ij}$ of degree~$J$
\begin{gather*}
\cHQJ  \equiv  \big\{P \in \pP[Q_{ij}] \,|\, P(\rho
Q_{ij}) = \rho^J\,P(Q_{ij}),\; \forall\, \rho\in\C  \big\} .
\end{gather*}
These are polynomials completely anti-holomorphic in the spinors~$z_i$ and of order~$2J$.

One can prove that these Hilbert spaces $\cHQJ$ are isomorphic to
the Hilbert space $\cHNJ$  of $N$-valent intertwiners with f\/ixed
total area $J$. To this purpose, we will construct the explicit
representation of the operators quantizing $M_{ij}$ and $Q_{ij}$ on
the spaces $\cHQJ$ and show that they match the actions of the
$\U(N)$ operators $E_{ij}$ and $F^\dag_{ij}$ which we described
earlier.
Our quantization relies on quantizing the $\bz_i$ as multiplication
operators while promoting $z_i$ to a~derivative operator
\begin{gather*}
\widehat{\bz}_i^a  \equiv  {\bz}_i^a \times ,\qquad
\widehat{z}_i^a  \equiv  \f{\pp}{\pp \bz_i^a},
\end{gather*}
which satisf\/ies the commutator $[\hat{z},\hat{\bz}]=1$ as expected
for the quantization of the classical bracket $\{z,\bz\}=i$. Then,
we quantize the matrix elements $M_{ij}$ and $Q_{ij}$ and the
closure constraints following this correspondence:
\begin{gather*}
\widehat{M}_{ij}  =  {\bz}_i^0\f{\pp}{\pp
\bz_j^0}+{\bz}_i^1\f{\pp}{\pp \bz_j^1} , \qquad
\widehat{Q}_{ij}  =
\bz_i^0\bz_j^1-\bz_i^1\bz_j^0  =  Q_{ij} , \\
\widehat{\bQ}_{ij}
 =  \f{\pp^2}{\pp \bz_i^0\pp \bz_j^1}-\f{\pp^2}{\pp \bz_i^1\pp
\bz_j^0} , \qquad
\widehat{\cC}_{ab}  =
\sum_k{\bz}_k^b\f{\pp}{\pp\bz_k^a}.
\end{gather*}
It is straightforward to check that the $\wcC_{ab}$ and the
$\wM_{ij}$ respectively form a $\u(2)$ and a $\u(N)$ Lie algebra, as
expected
\begin{gather*}
[\wcC_{ab},\wcC_{cd}] =
\delta_{ad}\wcC_{cb}-\delta_{cb}\wcC_{ad},\qquad
[\wM_{ij},\wM_{kl}] =
\delta_{kj}\wM_{il}-\delta_{il}\wM_{kj},\qquad
[\wcC_{ab},\wM_{ij}] = 0,
\end{gather*}
which amounts to multiply the Poisson bracket~\eqref{commC} and~\eqref{commM}  by $-i$.
Then, we f\/irst check the action of the closure constraints on
functions of the variables $Q_{ij}$:
\begin{gather*}
 \widehat{\vcC} Q_{ij} =  0 ,\qquad
\widehat{(\tr \cC)} Q_{ij} =  2Q_{ij},\\ \forall\,
P\in\cHQJ=\pP_J[Q_{ij}],\qquad
 \widehat{\vcC}\,P(Q_{ij}) = 0,\qquad
\widehat{(\tr \cC)}\,P(Q_{ij}) =  2J P(Q_{ij}),
\end{gather*}
so that our wavefunctions $P\in\cHQJ$ are $\SU(2)$-invariant (vanish
under the closure constraints) and are eigenvectors of the
$\tr \cC$-operator with eigenvalue~$2J$.

Second, we check that the operators $\wM$ and $\widehat{(\tr\,\cC)}$
satisfy the same quadratic constraints on the Hilbert space $\cHQJ$
(i.e.\ assuming that the operators acts on $\SU(2)$-invariant
functions vanishing under the closure constraints) that the
$\u(N)$-generators $E_{ij}$
\begin{gather*}
\widehat{(\tr \cC)}=\sum_k \wM_{kk},\qquad \sum_k \wM_{ik}\wM_{kj}
 =  \wM_{ij}\left(\f {\widehat{(\tr \cC)}}2 +N-2\right),
\end{gather*}
which allows us to get the value of the (quadratic) $\U(N)$-Casimir
operator on the space~$\cHQJ$:
\[
\sum_{ik}\wM_{ik}\wM_{ki} =  \widehat{(\tr \cC)}\left(\f
{\widehat{(\tr \cC)}}2 +N-2\right)  =  2J(J+N-2).
\]
Thus, we can safely conclude that this provides a proper
quantization of our spinors and $M$-variables, which matches exactly
with the  $\u(N)$-structure on the intertwiner space (with the exact
same ordering)
\begin{gather*}
\cHQJ \sim\cHNJ,\qquad \wM_{ij} = E_{ij},\qquad
\widehat{(\tr \cC)} = E.
\end{gather*}

Now, turning to the $\wQ_{ij}$-operators, it is straightforward to
check that they have the exact same action that the~$F^\dag_{ij}$
operators, they satisfy the same Lie algebra commutators~\eqref{commEF} and the same quadratic constraints
(\ref{constraint1})--(\ref{constraint3}). Clearly, the simple
multiplicative action of an operator~$\wQ_{ij}$ send a polynomial in~$\pP_J[Q_{ij}]$ to a polynomial in $\pP_{J+1}[Q_{ij}]$.
Reciprocally, the derivative action of~$\wbQ_{ij}$ decreases the
degree of the polynomials and maps~$\pP_{J+1}[Q_{ij}]$ onto~$\pP_J[Q_{ij}]$.

Finally, let us look at the scalar product on the whole space of
polynomials $\pP[Q_{ij}]$. In order to ensure the correct
Hermiticity relations for~$\wM_{ij}$ and $\wQ_{ij}$, $\wbQ_{ij}$, it
seems that we have a unique measure (up to a global factor)
\begin{gather*}
\forall\, \phi,\psi\in \pP[Q_{ij}],\qquad
\la\phi|\psi\ra  \equiv  \int \prod_id^4z_i e^{-\sum_i \la
z_i|z_i\ra}  \overline{\phi(Q_{ij})} \psi(Q_{ij}) .
\end{gather*}
Then it is easy to check that we have $\wM^\dag_{ij}=\wM_{ji}$ and
$\wQ^\dag_{ij}=\wbQ_{ij}$ as wanted.

It is easy to see that the spaces of homogeneous polynomials
$\pP_J[Q_{ij}]$ are  orthogonal with respect to this scalar product.
The quickest way to realize that this is true is to consider the
operator $\widehat{(\tr \cC)}$, which is Hermitian with respect to
this scalar product and takes dif\/ferent values on the spaces
$\pP_J[Q_{ij}]$ depending on the value of~$J$. Thus these spaces
$\pP_J[Q_{ij}]$ are orthogonal to each other.

This concludes our quantization procedure thus showing that the
intertwiner space for $N$ legs and f\/ixed total area $J=\sum_i j_i$
can be seen as the space of homogeneous polynomials in the $Q_{ij}$
variables  with degree $J$. This provides us with a description of
the intertwiners as wave-functions anti-holomorphic in the spinors
$z_i$ constrained by the closure conditions\footnote{It is also
possible to present an alternative construction~\cite{Borja:2010rc},
which can be considered as ``dual'' to the representation def\/ined
above. It is based on the coherent states for the oscillators, thus
recovering the framework of the $\U(N)$ coherent intertwiner states
introduced in \cite{Freidel:2010tt} and further developed in~\cite{Dupuis:2010iq}.}.

\subsubsection{Action principle}

It is possible to write an action principle for the previous Poisson
bracket structure in terms of the spinors. In order to be
consistent, we have to take into account the closure constraint, but
also the matching conditions coming from the gluing of several
intertwiners together. Furthermore, we propose an interaction term
(a Hamiltonian) for this model.

Once we know that the Hilbert space of LQG can be described as the
quantization of the phase space in terms of spinors (with the
construction explained above) \cite{Livine:2011gp}, it is
interesting to present explicitly the correspondence between the
standard formalism of loop (quantum) gravity and the spinor
formulation provided by the reconstruction of the $\SU(2)$ group
element~$g_\ell$ in terms of the spinors~\cite{Freidel:2010bw}.

Considering an link $\ell$ with the two spinors at each of its
end-nodes $z_{s(\ell),\ell}$ and $z_{t(\ell),\ell}$, there exists a unique
$\SU(2)$ group element mapping one onto the other. More precisely
\begin{gather*}
g_\ell \equiv  \f{|z_{s(\ell),\ell}]\la z_{t(\ell),\ell}|-|z_{s(\ell),\ell}\ra
[z_{t(\ell),\ell}|} {\sqrt{\la z_{s(\ell),\ell}|z_{s(\ell),\ell}\ra\la
z_{t(\ell),\ell}|z_{t(\ell),\ell}\ra}}
\end{gather*}
is uniquely f\/ixed by the following conditions
\begin{gather*}
g_\ell\,\f{|z_{t(\ell),\ell}\ra}{\sqrt{\la
z_{t(\ell),\ell}|z_{t(\ell),\ell}\ra}}=\f{|z_{s(\ell),\ell}]}{\sqrt{\la
z_{s(\ell),\ell}|z_{s(\ell),\ell}\ra}},\\ g_\ell\,\f{|z_{t(\ell),\ell}]}{\sqrt{\la
z_{t(\ell),\ell}|z_{t(\ell),\ell}\ra}}=-\f{|z_{s(\ell),\ell}\ra}{\sqrt{\la
z_{s(\ell),\ell}|z_{s(\ell),\ell}\ra}},\qquad g_\ell\in\SU(2),
\end{gather*}
thus sending the source normalized spinor onto the dual of the
target normalized spinor.

\begin{figure}[h]
\centering
\includegraphics[height=60mm]{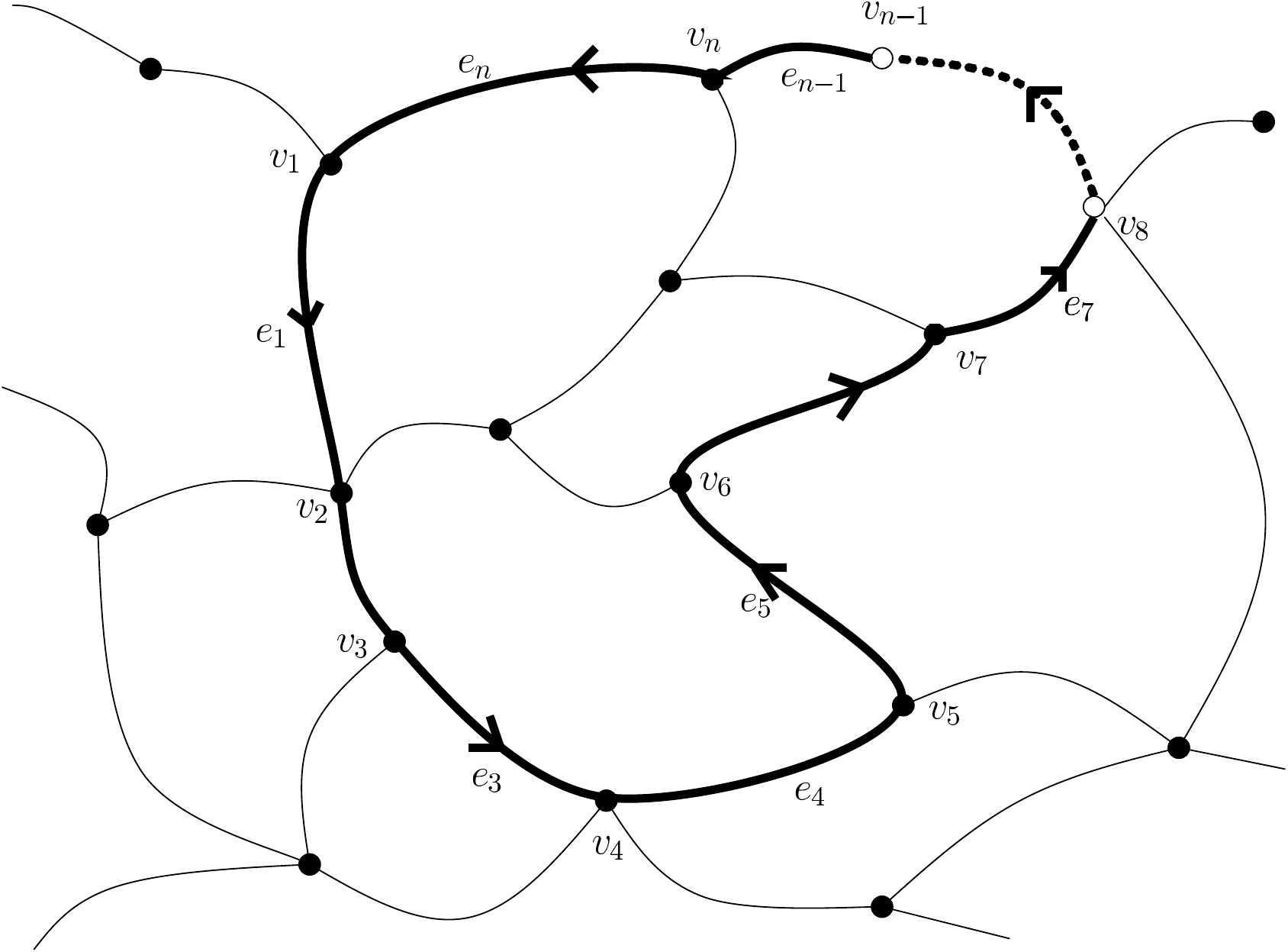}
\caption{The loop $\cL=\{\ell_1,\ell_2,\dots,\ell_n\}$ on the graph $\Gamma$.}\label{loop_fig}
\end{figure}

Starting from this point, it is possible to construct objects in
terms of the matrices~$M$ and~$Q$ that are~$\SU(2)$ invariant and
satisfying the matching conditions~\cite{Borja:2010rc}. The
expression of these objets (dubbed as ``generalized holonomies'' for
their close relation with the usual holonomy operators in LQG) is
\begin{gather*}
\cM_\cL^{\{r_i\}}  \equiv  \prod_i r_{i-1}r_i\bQ^i_{i,i-1} +
(1-r_{i-1})r_iM^i_{i-1,i} +  r_{i-1}(1-r_i)M^i_{i,i-1} \\
\phantom{\cM_\cL^{\{r_i\}}  \equiv}{} +
(1-r_{i-1})(1-r_i)Q^i_{i,i-1}
 =  \prod_i \la \varsigma^{r_{i-1}}
z_{v_i,e_{i-1}}\, |\, \varsigma^{1-r_i} z_{v_i,e_i}\ra
\end{gather*}
with $r_i=0,1$; and $M^v_{j,k}$, $Q^v_{j,k}$ the corresponding
operators for the node $v$ of a given loop of a~generic graph
$\Gamma$ (Fig.~\ref{loop_fig}) and acting on the links~$j$ and~$k$.
These objects will be the building blocks for the interaction term
of our model.

Using all these ingredients, we want to write an action principle
for this formalism. We should keep in mind that the spinnetwork
states on a given graph $\Gamma$ are $V$ intertwiner states~-- one at
each node~$v$~-- glued together along the links $e$ so that they
satisfy the matching conditions on each link. Consequently the phase
space consists with the spinors $z_{v,\ell}$ (where $e$ are links
attached to the node $v$, i.e.\ such that $v=s(\ell)$ or $v=t(\ell)$) which
we constrain by the closure conditions~$\vcC^v$ at each node $v$ and
the matching conditions on each link $e$. The corresponding action
reads
\begin{gather*}
S_0^{\Gamma}[z_{v,\ell}]  =  \int dt\,\sum_v \sum_{e|v\in\pp e}
\left(-i\la z_{v,\ell}|\pp_n z_{v,\ell}\ra +\la z_{v,\ell}|\Lambda_v|
z_{v,\ell}\ra\right)\\
\phantom{S_0^{\Gamma}[z_{v,\ell}]  =}{}
 +\sum_\ell \rho_\ell\left(
 \la z_{s(\ell),\ell}|z_{s(\ell),\ell}\ra-\la z_{t(\ell),\ell}|z_{t(\ell),\ell}\ra
 \right),
\end{gather*}
where the $2\times 2$ Lagrange multipliers  $\Lambda_v$ satisfying
$\tr \Lambda_v=0$ impose the closure constraints and the Lagrange
multipliers $\rho_\ell\in\R$ impose the matching conditions. All the
constraints are f\/irst class, they generate~$\SU(2)$ transformations
at each node and~$\U(1)$ transformations on each link~$e$.

We can analogously describe this system parameterized by $N_v\times
N_v$ unitary matrices $U^v$ and the parameters $\lambda_v$. The
matrix elements $U^v_{ef}$ refer to pairs of links $e$, $f$ attached to
the node~$v$. As it was mention before, the closure conditions are
automatically encoded in the requirement that the matrices $U^v$ are
unitary. Of course, we still have to impose the matching conditions
$M^{s(\ell)}_{ee}-M^{t(\ell)}_{ee}=0$ on each link $e$ where the matrices
$M^v=\lambda_v\,U^v\Delta U^v{}^{-1}$ are functions of both~$\lambda_v$ and $U^v$. So in this case, the action\footnote{This action is invariant under the action of~$\SU(2)\times\U(N_v-2)$ at every node, which reduces the number of
degrees of freedom of the matrices $U^v$ to the spinors~$z_{v,\ell}$
which are actually the two f\/irst columns of those matrices.}
reads
\begin{gather*}
S_0^{\Gamma}[\lambda_v,U^v] =  \int dt
\sum_v\big({-}i \lambda_v \tr U^v\Delta \pp_n {U^v}^\dag  - \tr
\Theta_v (U^v{U^v}^\dag-\id) \big) +\sum_\ell
\rho_\ell(M^{s(\ell)}_{ee}-M^{t(\ell)}_{ee}),
\end{gather*}
where the $\rho_\ell$ impose the matching conditions while the
$N_v\times N_v$ matrices~$\Theta_v$ are the Lagrange multipliers for
the unitarity of the matrices~$U^v$.

This free action describes the classical kinematics of spinnetworks
on the graph $\Gamma$. Now, we are going to add interaction terms to
this action. Such interaction terms are built with the generalized
holonomy observables $\cM_\cL^{\{r_i\}}$. With this construction,
the closure and matching conditions are trivially satisf\/ied. Our
proposal for a classical action for spinnetworks with nontrivial
dynamics is thus
\begin{gather*}
S_{\gamma_\cL^{\{r_i\}}}^{\Gamma}=S_0^{\Gamma} +  \int
dt \sum_{\cL,\{r_i\}}
\gamma_\cL^{\{r_i\}} \cM_\cL^{\{r_i\}},
\end{gather*}
where the $\gamma_\cL^{\{r_i\}}$ are the coupling constants giving
the relative weight of each generalized holonomy in the full
Hamiltonian. We will study in more detail this classical action
principle in the specif\/ic case of the 2-node graph in the following
section.

\subsubsection{Ef\/fective dynamics for the 2-node graph}

Let us particularize to the 2-node graph the action principle
proposed before for a general graph. Then, the action for this
model, including a general interaction term is
\begin{gather}
S[\Ua,\Ub,\lambda]  \equiv  S_0[\Ua,\Ub,\lambda]  + \int dt
\sum_{i,j} \big[ \gamma^+_{ij} Q^\alpha_{ij}Q^\beta_{ij} +
\gamma^-_{ij} \bQ^\alpha_{ij}\bQ^\beta_{ij}
+\gamma^0_{ij} M^\alpha_{ij}M^\beta_{ij} \big],
\label{genericaction}
\end{gather}
with
\begin{gather*}
S_0[\Ua,\Ub,\lambda]\equiv\int dt\bigg({-}i
\lambda\big[\tr\Ua\Delta\pp_n\Ua{}^\dag+\tr\Ub\Delta\pp_n\Ub{}^\dag\big]\\
\phantom{S_0[\Ua,\Ub,\lambda]\equiv}{}
+\sum_i \rho_i\big[
(\Ua\Delta\Ua{}^\dag)_{ii}-(\Ub\Delta\Ub{}^\dag)_{ii}
\big]\bigg),
\end{gather*}
where $\lambda\equiv\lambda_\alpha=\lambda_\beta$, due to the
matching conditions, and the $\gamma$'s are coupling constants
satisfying
\[
\gamma^- =\overline{\gamma^+},\qquad \gamma^0=(\gamma^0)^\dag
\]
in order to have a real Hamiltonian.

At this point one can look for the classical counterpart of the
quantum Hamiltonian for the homogeneous and isotropic sector
imposing a global $\U(N)$ symmetry. After this, the action depends
just on two conjugated variables $\lambda\equiv\tr M/2$ and $\phi$.
Due to the symmetry reduction and the matching conditions, $\phi$
relates the unitary matrix (or spinor) in the node~$\alpha$ with the
one at~$\beta$: $\Ua = e^{i\phi}  \overline{\Ub}$. Finally, the
expression for the action for the reduced sector is
\begin{gather*}
S_{\rm inv}[\lambda,\phi]  =  -2 \int dt \big(\lambda \pp_n \phi
-\lambda^2\big(\gamma^0-\gamma^+e^{2i\phi}
-\gamma^-e^{-2i\phi}\big)\big), 
\end{gather*}
with the Hamiltonian $H=\lambda^2(\gamma^0-2\gamma\cos(2\phi))$.

This Hamiltonian corresponds with the quantum Hamiltonian $H$ that
we have considered before.  As we did also there, we can introduce
the renormalized Hamiltonian
\begin{gather*}
\hh \equiv  \f1\lambda H=\lambda(\gamma^0-2\gamma\cos(2\phi)),
\end{gather*}
that is still $\SU(2)$ and $\U(N)$ invariant.

The equations of motion coming from the new Hamiltonian $\hh$ are
simply given by
\begin{gather*}
\partial_n\phi = \gamma^0-2\gamma\cos(2\phi),\qquad
\partial_n\lambda = -4\gamma\lambda\sin(2\phi) .
\end{gather*}
We can solve exactly these dif\/ferential equations. First we solve
for $\phi(t)$ analytically and then the following expression for
$\lambda$ in terms of $\phi$ solves the equations of motions
\begin{gather}
\label{conicparam}
\lambda=\frac{\eps}{\gamma^0-2\gamma\cos(2\phi)} ,
\end{gather}
where $\eps=\pm$ is a global sign. Let us point out that the
equation of motion for~$\lambda$ only determines it up a global
numerical factor. Then we should remember that~$\lambda$ is the
total area and we always constrain it to be positive. Moreover this
is the equation of a conic with radial coordinate given by~$\lambda$, polar coordinate~$2\phi$ and eccentricity~$2\gamma/\gamma_0$.

The solutions\footnote{We have chosen the most convenient constants of
integration due to the fact that this constants are just
translations in the temporal variable}  for $\phi(t)$,
depending on the dif\/ferent values for the parameters $\gamma^0$ and
$\gamma$ are

{\bf elliptic region} $(|\gamma^0|>2|\gamma|)$:
\begin{subequations}\label{regions}
\begin{gather}
\phi(t)=-\arctan\left(\frac{(2\gamma-\gamma^0)\tan\left(t\sqrt{(\gamma^0)^2-4\gamma^2}\right)} {\sqrt{(\gamma^0)^2-4\gamma^2}}\right),
\end{gather}

{\bf hyperbolic region} \ $(|\gamma^0|<2|\gamma|)$:
\begin{gather}
\phi(t)=-\arctan\left(\frac
{\sqrt{4\gamma^2-(\gamma^0)^2}}{(2\gamma+\gamma^0)\tanh\left(t\sqrt{4\gamma^2-(\gamma^0)^2}\right)}\right),
\end{gather}

{\bf parabolic region I} $(\gamma^0=2\gamma)$:
\begin{gather}
\phi(t)=-\arctan\left(\frac{1}
{4\gamma t}\right) ,
\end{gather}

{\bf parabolic region~II} $(\gamma^0=-2\gamma)$:
\begin{gather}
\phi(t)=-\arctan\left(4\gamma t\right).
\end{gather}
\end{subequations}

\begin{figure}[h]
\centering
\includegraphics[height=100mm]{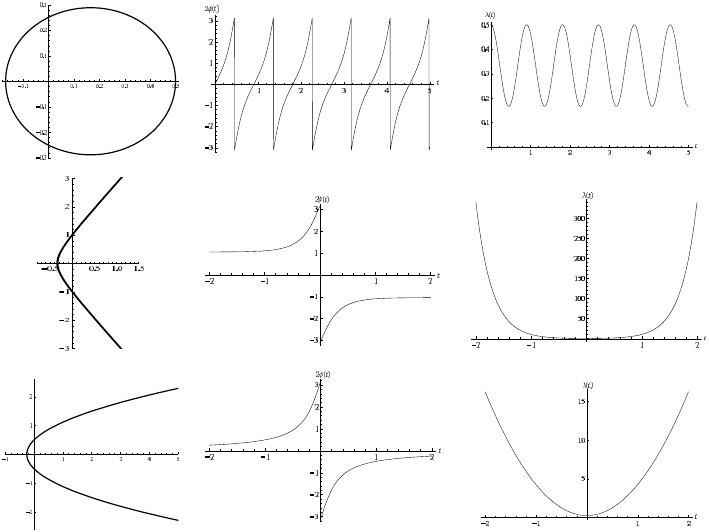}
\caption{We plot the behavior of $\phi(t)$ and $\lambda(t)$ (given
by the equations~\eqref{conicparam} and~\eqref{regions}) in the three
dif\/ferent regimes for $\gamma=1$ and respectively $\gamma^0=4$
(elliptic regime), $\gamma^0=1$ (hyperbolic regime) and f\/inally
$\gamma^0=2$ (parabolic regime). In the f\/irst column, we give the
polar plots constructed by taking as polar coordinates $(2
\phi,\lambda(\phi))$. The second column gives for $\phi(t)$ and the
third one  $\lambda(t)$. We observe in those plots the periodical
behavior of $\lambda$ (interpreted as the total area of the model)
as a function of time in the elliptic case and a behavior  analogous
to a cosmological big bounce in the other two cases.} \label{conics}
\end{figure}

Let us give a brief description of these solutions (Fig.~\ref{conics}). First, we notice that we get the same solution
$\lambda(t)$ for the two cases~I and~II of the parabolic region by
taking $\eps=+$ in case~I and $\eps=-$ in case~II. In the elliptic
case, we have a system in which the area $\lambda$ has an
oscillatory behavior. While in the other two regimes the area
shrinks under evolution, reaches a~minimum value and then increases
until inf\/inity. As it was pointed out in~\cite{Borja:2010gn,Borja:2010rc}, the quantum Hamiltonian of this
2-node model is mathematically analogous to the gravitational part
of the Hamiltonian in LQC. Following this analogy, we can interpret
the results obtained here as the classical analogous of the quantum
big bounce found in LQC.

At this point, it would be very interesting to go beyond the
$\U(N)$-invariant sector. The action~\eqref{genericaction} def\/ines the
full classical kinematics and dynamics of spin network states on the
2-node graph. It is a nontrivial matrix model def\/ined in terms of
the unitary matrices~$\Ua$ and~$\Ub$ and with quartic interaction
terms. Even if we still choose a~$\U(N)$-invariant Hamiltonian of
the the type $\gamma^+ \tr  Q^\alpha Q^\beta +
\gamma^- \tr \bQ^\alpha\bQ^\beta +\gamma^0 \tr M^\alpha
{}^tM^\beta$, this will nevertheless induce nontrivial dynamics for
the matrices~$\Ua$ and~$\Ub$. It would be very interesting to follow
this line of research and study what kind of anisotropy does this
model describe in the context of loop cosmology.

Finally, let us point out some other possible lines of research on these topics.  Although the results on the $2$-node graph are compelling due to the relation with cosmology, there are some limitations. Up to now, it has not been possible to go beyond the homogeneous sector. The relation with the improved dynamics in LQC and the introduction of matter are still missing.  It would be interesting to generalize the methods presented here to more complicated graphs in order to test the truncation of LQG to a f\/ixed graph. We point out two such generalizations. One is the work with the graph with $3+N$ nodes in \cite{Borja:2010gn}.
This graph could allow us, for instance, to study rotations or black-hole radiation processes.  An interesting  generalization could be to consider a graph with an inf\/inite number of nodes in order to study a continuum limit. These models can be generalized in numerous directions and we think that they can contribute to the understanding of the framework and shed light on fundamental problems in LQG and spinfoams.

\section{The covariant few-node model}  \label{section4}

So far we have presented the use of the 2-nodes graph in the canonical theory, seeing how this can be used as a truncation of the full theory to a f\/inite number of degrees of freedom, and we have seen dif\/ferent techniques to impose homogeneity and isotropy on this system in order to obtain the description of a FLRW universe.
On the other hand, in our description we can not break free from the ambiguities in the def\/inition of the dynamics: there is no agreement about the form of the Hamiltonian constraints used. We hope that the works presented in this review could add new insights to this question.

The dynamics of Loop Quantum Gravity, however, admits also a covariant formulation in terms of transition amplitudes, which appears to be far less subject to ambiguities. The fundamental object is a general covariant path integral, interpreted as a \emph{sum over geometries}. In this section we review calculation of the transition amplitude between homogeneous isotropic states on a regular graph. This calculation was f\/irst performed in the context of the 2-node model~\cite{Bianchi:2010zs} and extended to the case with a cosmological constant in~\cite{Bianchi:2011ym}, but was later extended to an arbitrary regular graph in~\cite{Vidotto:2011qa}.  Here we give directly the general case with a cosmological constant on an arbitrary regular graph.

\subsection{Brief introduction to spinfoam theory}

The transition amplitudes are obtained by summing over all the possible spinfoams.
A \emph{spinfoam} is a 4-dimensional simplicial 2-complex $\cal C$
colored with \emph{spins} $j_f$ and intertwiners $i_e$, associated, respectively,
to the faces $f$ and the edges~$e$, i.e.\ the 3-cells. This is the object that encodes the quantum geometry (Penrose's spin-geometry theorem).
Here we take the sum over the coloring $j_f$ and $i_\ell$,
the product of ``face amplitude'' $\prod_f d(j_f) $ and a product of vertex amplitudes $\prod_v A_v(j_\ell,i_v)$, that reads
\cite{Engle:2007wy,Engle:2007qf,Engle:2007uq,Freidel:2007py,Kaminski:2009fm,Livine:2007hc}:
\begin{gather}\label{zamp}
   Z_{\cal C}=\sum_{j_f,i_e} \prod_f d(j_f)   \prod_v A_v(j_f,\vol_e).
\end{gather}

The vertex amplitude $A_v(j_f,\vol_e)=\langle j_f,\vol_e| A_v\rangle$ is written in a basis of intertwiners that diagonalizes the volume, and we indicate with $\vol_e$ the corresponding quantum number, which we take to be the eigenvalue (for simplicity of notation we disregard the eventual degeneracy). Thus the vertex amplitude is a function of the spins~$j_f$ and of the intertwiners adjacent to the vertex~$v$.

We can include a positive cosmological constant by considering a simple modif\/ication of~\eqref{zamp}  based on the form of the cosmological-constant term in the Hamiltonian constraint.
 In the canonical theory, the cosmological constant appears as an additive term to the gravitational Hamiltonian constraint, which multiples the 3-volume element. When deriving a path integral formulation of quantum theory \`a la Feynman by inserting resolutions of unity into the evolution operator, a potential term appears simply as a multiplicative exponential, because the potential is diagonal in the position basis. The cosmological constant term is diagonal in the spin-intertwiner basis.  It is therefore possible to insert the cosmological constant ``potential'' as a multiplicative term along the spinfoam evolution, that is in between 4-cells, which is to say on 3-cells\footnote{The boundary state of each cell is written in the time gauge~\cite{Rovelli:2011ly}.}. The coupling is therefore very simple, and consists in weighting edge amplitudes with an exponential term which depends on the volume and the cosmological constant.
Therefore we obtain
\begin{gather}
   Z_{\cal C}=\sum_{j_f,\vol_e} \prod_f (2j+1)  \prod_e e^{i\lambda \vol_e} \prod_v A_v(j_f,\vol_e) ,
\label{cosmo.ampli}
\end{gather}
where $\lambda$ is related to the cosmological constant $\Lambda$ and
eigenvalue of the volume $\vol_e$ associated to an edge $e$. The amplitude is written in a basis of intertwiners that diagonalizes the volume, so that the term with $\vol_e$ in the exponential is well def\/ined.

Incorporating this term into the covariant dynamics of Loop Quantum Gravity (see~\mbox{\cite{Rovelli:2010wq, Rovelli:2010vv}} and references therein) is important in order to check the semiclassical limit. Einstein equation admits only the trivial f\/lat solution in absence of matter for $\Lambda=0$. Recovering f\/lat space is interesting, but is still weak evidence for the full classical limit.  Since at the moment the coupling of matter in spinfoam is not yet completely understood, the inclusion of the cosmological constant became essential to check the good semiclassical limit of the spinfoam theory beyond the trivial f\/lat solution.

\subsubsection{Coherent states}

We want to study the semiclassical behavior of this transition amplitude. In  in the Hilbert space ${\cal H}_\Gamma$
it is possible to def\/ine an overcomplete basis of semiclassical states, obtained as a~superposition of spinnetwork states. These are coherent states, functions of $\SU(2)$ and  labelled by a $\SL(2,\C)$ element $H_\ell$ for each link. They take the form\footnote{As shown in  \cite{Bianchi:2009ky}, these states: (i)~are the basis of the holomorphic representation~\cite{Ashtekar:1994nx,Bianchi:2010ys}, (ii)~are a special case of Thiemann's complexif\/ier's coherent states~\cite{Bahr:2007xa,Bahr:2007xn,Flori:2009rw,Flori:2008nw, Sahlmann:2001nv,Thiemann:2002vj,Thiemann:2000bw,Thiemann:2000ca,Thiemann:2000bx, Thiemann:2000by},
(iii)~induce Speziale--Livine coherent tetrahedra \cite{Conrady:2009px,Freidel:2009nu, Livine:2007hc} on the nodes, and (iv) are equal to the Freidel--Speziale coherent states \cite{Freidel:2010uq,Freidel:2010bw} for large spins.}
\cite{Bianchi:2009ky,Bianchi:2010ys}
\begin{gather}
\psi_{H_\ell}(h_\ell) = \int_{\SU(2)^N} dg_n
    \prod_{l\in \Gamma} K_{t}\big(  g_{s(\ell)}  h_\ell   g^{-1}_{t(\ell)}   H_\ell ^{-1}\big).
 \label{states}
 \end{gather}
They are def\/ined by an integral on $\SU(2)$, so that the stases are gauge invariant, and  by the \emph{heat kernel} $K_t$ on $\SU(2)$ $\big(h_\ell\in \SU(2)\big)$, analytically continued to $\SL(2,\C)$. This is a function concentrated on the origin of the group, with a spread of order~$1/t$ in $j$. Its explicit form is\footnote{We choose a parameter $t$ with the dimension of an inverse action, and put $\hbar$ explicitly in the def\/inition of the coherent states, in order to emphasize the fact that the small $t$ limit is the classical limit, and to keep track of the corresponding dependence on~$\hbar$. The factor 2 is for later convenience.}
\begin{gather*}
K_t(h) = \sum_j (2j+1)  e^{-2 t\hbar\, j(j+1)}   \tr[D^j(h)],
\end{gather*}
where $D^j(h)$ is the Wigner matrix of the spin-$j$ representation of $\SU(2)$.
The states \eqref{states} are gauge-invariant semiclassical wave packets.  The integral in~\eqref{states} projects (``group averages'') on the gauge invariant states. If~$H_\ell$ is in the~$\SU(2)$ subgroup of $\SL(2,\C)$, the heat kernel peaks each~$h_\ell$ on~$H_\ell$. The extension of~$H_\ell$ to~$\SL(2,\C)$ has the same ef\/fect as taking a gaussian function $\psi(x)=e^{(x-z_o)^2/2}\sim e^{(x-x_o)^2/2} e^{ip_ox}$ for a complex $z_o=x_o+ip_o$; that is, it adds a phase which peaks the states on a value of the variable conjugate to~$h_\ell$. Thus, the states~\eqref{states} are peaked on the variables~$h_\ell$ as well as on their conjugate momenta.

We can decompose each $\SL(2,\C)$ label in the form
\begin{gather*}
 H_\ell=
 D^{ (j)}(R_{\vec n_{s(\ell)}})
e^{-iz_\ell\f{\sigma_3}2}
D^{ (j)}\big(R_{\vec n_{t(\ell)}}^{-1}\big),
 \end{gather*}
where $R_{\vec n}\in \SU(2)$ is the rotation matrix that rotates the unit vector pointing in the $(0,0,1)$ direction into the unit vector $\vec {n}$, and $D^{(j)}(R_{\vec n_s})$ is its representation~$j$.
$\vec \sigma=\{\sigma_i\}$, $i=1,2,3$ are the Pauli matrices.

There is a compelling geometrical interpretation for the $(\vec n_s, {\vec n}_t, \xi, \eta)$ labels of each link~\cite{Freidel:2010uq,Magliaro:2010vn,Rovelli:2010km}.  The two vectors~$\vec n_s$ and~$\vec n_t$ represent the normals to the face~$\ell$, in the two polyhedra bounded by this face. The complex number~$z_\ell$ codes the intrinsic and the extrinsic geometry at the face. More precisely the imaginary part of~$z_\ell$ is proportional to the area of the face of the cellular decomposition dual to the link~$\ell$. The real part of~$z_\ell$ is determined by the holonomy of the Ashtekar connection along the link~\cite{Rovelli:2010km}.
For general states, the interpretation extends to a simple generalization of Regge geometries, that Freidel and Speziale have baptized ``twisted geometries''~\cite{Freidel:2010uq}.

These state, that we use to concretely compute the transition amplitude,
should be interpreted%
\footnote{They can also be viewed as describing quantum space at some given coordinate time, but this interpretation is less covariant.}
as describing the quantum space  surrounding a given 4-dimensional f\/inite region of spacetime.  We talk therefore of ``boundary states'', that can be thought here as ``in'' and ``out'' states in the transition amplitude.

\subsubsection{Vertex amplitude}

The transition amplitude (\ref{zamp}), (\ref{cosmo.ampli}) can be expressed as a vertex expansion. The f\/irst nontrivial term of this expansion involves just a single vertex. In what follows we concentrate on the evaluation of the transition amplitude in the f\/irst order of the vertex expansion. Therefore
the vertex amplitudes $A_v(H_\ell)$, one for each vertex~$v$ in the bulk of the spinfoam, become for us just~$A(H_\ell)$.
\begin{figure}[t]
\centering
\includegraphics{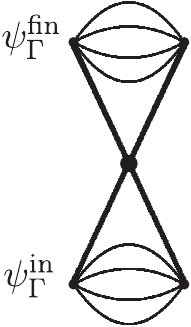}

\caption{Transition amplitude between two states def\/ined on a ``dipole'' graph. We consider only the f\/irst order in the vertex expansion, i.e.\ there is only one vertex in the bulk (spinfoam edges are drawn with thicker lines).\label{edgy}}
\end{figure}
We
 evaluate it in the basis of the coherent states
  \cite{Barrett:2009bs,Bianchi:2011ta,Bianchi:2010ys,Engle:2008ev, Pereira:2007nh}, so that the expression $
\langle A | \psi_{H_\ell}  \rangle
=W(H_\ell)
$ reads
\begin{gather}
W(H_{\ell})=  \int_{\SL(2,\C)}  \prod_{n=1}^{N-1} \d G_n
\prod_{\ell=1}^ {L}
\sum_{j_\ell} {\scriptstyle (2j_\ell+1)}e^{ -2t\hbar j_\ell(j_\ell+1)}
e^{i\lambda\vol_e}\nonumber\\
\hphantom{W(H_{\ell})=}{} \times
 \tr
\big[
D^{ (j_\ell)}(H_\ell)
Y^\dagger  D^{ (\gamma j_\ell, j_\ell)}(G_\ell) Y
 \big].\label{ampi}
\end{gather}
The amplitude is Lorentzian, with an integration over the $\SL(2,\C)$ elements $G_n$ associated to the edges (see Fig.~\ref{edgy}). Notice that the integration is over all the $G_n$ but one, in order to avoid a redundancy that makes the amplitude diverge~\cite{Engle:2008ev}.
We take the product over each link $\ell$ and the sum over the coloring of these links, i.e.\ on the spins $j_\ell$.
$
D^{ (j)}(H_\ell)$ is simply
$
D^{ (j)}(R_{\vec n_{s(\ell)}})
D^{ (j)}(e^{-iz_\ell\f{\sigma_3}2})
D^{ (j)}(R_{\vec n_{t(\ell)}}^{-1}) .
$
$G_\ell=G_{s(\ell)}G_{t(\ell)}^{-1}$ is the product of the $\SL(2,\C)$ group elements at the source and target nodes, extremals of each oriented link $\ell$, and $D^{ (\gamma j_\ell, j_\ell)}(G_\ell)$ is its representation matrix.
Finally,
$Y$ is a map from the representation $(j)$ of $\SU(2)$ to the representation $(\gamma j_\ell, j_\ell)$ of $\SL(2,\C)$.
The f\/irst has has dimension $2j+1$ while the second has inf\/inite dimension. These matrices with dif\/ferent dimensions are glued by the map $Y$. In other words
\begin{alignat*}{5}
& Y: \  &&  \H^{(j)} && \longrightarrow \ &&   \H^{(j,\gamma j)} & \\
& &&       \ket{j,m} &&&&  \ket{(j,\gamma j);j,m}&
\end{alignat*}
whose matrix elements are given by
$
 \bek{(j,\gamma j);j',m'}Y{j,m} = \delta_{p,\gamma j}\delta_{kj}\delta_{jj'}\delta_{mm'}$.

In the base of the coherent states the amplitude takes the convenient form~\eqref{ampi}
 that we exploit for the calculation in cosmology.

\subsection{Homogeneous and isotropic geometry}

We want to evaluate the vertex amplitude \eqref{ampi} for the homogeneous and isotropic case \cite{Bianchi:2011ym,Bianchi:2010zs,Vidotto:2011qa,Vidotto:2010kw}. This corresponds to restrict the study to regular graphs, i.e.\ graphs where the distribution of the degrees
at the nodes is uniform (this condition is trivially satisfy when we work with the dipole).
The requirement of homogeneity and isotropy
f\/ixes $\vec n_s$, $\vec n_n$ as the normals to the faces of the geometrically regular cellular decomposition dual to
the graph, and implies that all the $z_\ell$ elements in $H_\ell$ are equal: $z_\ell=z$.  Furthermore, on a
homogeneous isotropic space the real part of $z$ is the sum of two terms~\cite{Magliaro:2010qz}
$
\textrm{Re}\,  z = \theta (\gamma K + \Gamma),
$
where $K$  and $ \Gamma $ are the scalar coef\/f\/icients of respectively the extrinsic curvature  and the spin connection, that enter in the def\/inition of the Ashtekar--Barbero connection written in the homogeneous gauge. On a compact space, $\Gamma=1$, and~$\theta$ and is the angle between two 4d normals of the two adjacent polyhedra (the isotropy requires that this is the same for every coupe of normals) and~$K$ is proportional to the time derivative of the scale factor.
Finally, all the cells are equal and we can write~$\vol_e$ in the cosmological constant term as the volume~$\vol_o$ of a regular cell with faces having unit area, times~$j^{\frac32}$.

With these assumptions, any homogeneous isotropic coherent state on any regular graph is described by a single complex variable $z$, whose imaginary part is proportional to the area of each regular face of the cellular decomposition (and it can be put in correspondence with the total volume) and whose real part is related to the extrinsic curvature~\cite{Rovelli:2010km}.  We denote $\psi_{H_\ell(z)}$
this state, and $\psi_{H_\ell(z,z')}=\psi_{H_\ell(z)}\otimes \psi_{H_\ell(z')}$ the state on two copies of the regular graph,
obtained tensoring the ``in'' and ``out'' homogeneous isotropic states.

Before studying further our transition amplitude, let us consider the vertex expansion. We consider the classical Hamilton function of a homogeneous isotropic cosmology: this results in a dif\/ference between two boundary terms. With the cosmological constant $\Lambda$ it gives
\begin{gather*}
S_H=
\int {\mathrm{d}}t  \left(a\dot a^2 + \frac\Lambda3 a^3\right)\Big|_{\dot a =\pm \sqrt{\frac\Lambda3} a}=
\frac23 \sqrt{\frac\Lambda3} \big(a^3_{\fin}-a^3_{\ini}\big),
\end{gather*}
where  $a$ is the scale factor and $\dot a$ its time derivative.
 Therefore at the f\/irst order in $\hbar$ the quantum transition amplitude factorizes:
\begin{gather*}
  W(a_{\fin},a_{\ini})=e^{\frac i \hbar  S_H(a_{\fin},a_{\ini})}=W(a_{\fin})\overline{W(a_{\ini})}  .
  \end{gather*}
The same happens for the spinfoam amplitude
\begin{gather*}\label{factoriz}
\langle W |\psi_{H_\ell(z_{\fin},z_{\ini})}\rangle=W(z_{\fin},z_{\ini}) = W(z_{\fin})  \overline {W(z_{\ini})}
\end{gather*}
with $
W(z)\equiv W(H_\ell)$, where now~\eqref{ampi} edpends only on a single $z$ trough $H_\ell(z)$.

In \cite{Hellmann:2011jn}, Frank Hellmann points out that the factorization survives also beyond the classical (large distance) limit when we restrict to the one-vertex approximation of the amplitude, and observes that this factorization can be reinterpreted as the amplitude to go from the initial state to nothing and from nothing to the f\/inal state, namely as the contribution of a disconnected spacetime topology to the total transition amplitude.  This does not preclude assigning a~standard cosmological interpretation to the $W(z_{\fin},z_{\ini}) = W(z_{\fin})  \overline {W(z_{\ini})}$ amplitude: it is only a~consequence of the particularly simple dynamics of the classical symmetric system considered above, where the equations of motion determine a relation between~$a$ and its conjugate momentum~$p_a$ which is \emph{independent} from earlier values of the two. On the other hand, this observation is quite interesting because it allows each single term~$W(z)$ to be interpreted as the Hartle--Hawking ``wave function of the universe'' determined by a \emph{no-boundary} initial condition~\cite{Hartle:1983ai}. At the f\/irst order in the expansion, we can therefore  study $W(z)$ instead of $W(z_{\fin},z_{\ini})$ and interpret it as the ``wave function of the universe''.

We are interested in this quantity in the large volume limit, that correspond to take the imaginary part of $z$ is large. Let us consider separately the real and the imaginary part of~$z$.

When the imaginary part of $z$ is large we f\/ind
that the Wigner matrix in the trace  gives
\begin{gather*}
D^{ (j)}\big(e^{-iz\f{\sigma_3}2}\big)=\sum_m e^{-iz m}\, |m\rangle\langle m|.
\end{gather*}
For
Im\,$z \gg 1$ (large area) in this sum the term $m=j$ dominates, therefore
\begin{gather*}
D^{ (j)}\big(e^{-iz\f{\sigma_3}2}\big)\approx e^{iz j}\, |j \rangle\langle j|,
\end{gather*}
where $|j\rangle$ is the eigenstate of $L_3$ with maximum eigenvalue $m=j$ in the representation~$j$.
Inserting this result into~\eqref{ampi} we obtain
\begin{gather*}
W(z)= \int
\prod_{n=1}^{N-1} \d G_n
\prod_{\ell=1}^ {L}
\sum_{j_\ell}  (2j_\ell+1)  e^{ -2t\hbar j_\ell(j_\ell+1)
-i\lambda \vol_o j^{\frac32}
{ - i z_\ell j_\ell} }
\\
\hphantom{W(z)=}{}\times
\langle j_\ell|
  D^{ (j_\ell)}\big(R^{-1}_{\vec n_n}\big)
Y^\dagger  D^{ (\gamma j_\ell, j_\ell)}(G_\ell) Y
{ D^{ (j_\ell)}(R_{\vec n_s})}
 |j_\ell\rangle.
\end{gather*}
The action of the matrix $D^{ (j_{\ell})}(R_{\vec n_n})$ on the highest weights
states is precisely the def\/inition of the coherent states $|\vec n\rangle$, so we can write
\begin{gather}
W(z)= \int
\prod_{n=1}^{N-1} \d G_n
\prod_{\ell=1}^ {L}
\sum_{j_\ell}  (2j_\ell+1) e^{ -2t\hbar j_\ell(j_\ell+1)
-i\lambda \vol_o j^{\frac32}
{ - i z_\ell j_\ell} }\nonumber\\
\phantom{W(z)=}{}\times
  \bra{\vec n_{t(\ell)}}
Y^\dagger  D^{ (\gamma j_{\ell}, j_{\ell})}(G_\ell) Y
\ket{\vec n_{s(\ell)}} .\label{Wtostart}
\end{gather}

We can now study the  $\SL(2,\C)$ integral in \eqref{Wtostart} (without f\/ixing the $j$). Let us rewrite the previous expression as
\begin{gather}
W(z)=
\sum_{\{j_\ell\}}  \prod_{\ell=1}^ {L}    (2j_\ell+1)  e^{ -2t\hbar j_\ell(j_\ell+1)
-i\lambda \vol_o j^{\frac32}
{ - i zj_\ell} }\nonumber\\
\hphantom{W(z)=}{}\times
 \int    \prod_{n=1}^{N-1} \d G_n
\prod_{\ell=1}^ {L}
  \bra{\vec n_{t(\ell)}}
 Y^\dagger  D^{ (\gamma j_{\ell}, j_{\ell})}(G_\ell) Y
\ket{\vec n_{s(\ell)}}.\label{Wsplit}
\end{gather}
Since the gaussian sums in the f\/irst line peak the $j_\ell$'s over large values, the integral in the second line
can be computed in the large spin regime, where it can be evaluated using saddle point methods.  The computation of the integral in~\eqref{Wsplit} can be written in a spinor base, as the one introduced in~\cite{Bianchi:2011ta} and gives
\begin{gather} \label{integrale}
 \int
 \prod_{n=1}^{N-1} \d G_n
\prod_{\ell=1}^ {L}
 \bra{n_{s(\ell)}}
Y^\dagger  D^{ (\gamma j_{\ell}, j_{\ell})}(G_\ell) Y
\ket{n_{t(\ell)}}
=
\mathrm{H}
 \prod_{\ell=1}^ {L}
e^{-\frac12 i{ j_\ell} \theta
},
\end{gather}
where H is the Hessian of the logarithm of the integrand in~\eqref{integrale} \cite{Bianchi:2011ta} and~$\theta$
is a constant determined by the normals on the faces: it is the \emph{intrinsic} curvature
on the faces, coming from the spin connection in the Ashtekar connection.
We can def\/ine a new variable $\tilde z:= z-\theta$, so that the real part of $\tilde z$ is exactly the extrinsic curvature.

We can now compute the sum that appears in the amplitude
\begin{gather} \label{www}
W(z)=
\sum_{\{j\}} \mathrm{H}   \prod_{\ell=1}^ {L}   (2j_\ell+1)
e^{ -2t\hbar j_\ell(j_\ell+1)
-i\lambda \vol_o j^{\frac32}
  - i \tilde z j_\ell}
\end{gather}
by approximating it with a Gaussian integral peaked on  $j_\ell\sim j_o$.
We expand around  $j_0$  so that the new term is
$
     i\lambda \vol_o j^{\frac32}\sim  i\lambda \vol_o j_o^{\frac32}+
     \fracs32 i\lambda \vol_o j_o^{\frac12}\delta j.
$
The f\/irst term is a constant that can be reabsorbed in the normalization and the second contributes to
the phase.
The value of the peak of the gaussian $j_o$ is determined by the stationary point where the real part of the exponent in~\eqref{www} vanishes.
This gives a condition on the imaginary part of $\tilde z$ (associated to the area), that for large ($ j\gg1$) is
 \begin{gather}\label{Re}
 j_o\sim \textrm{Im}\,\tilde z/4t\hbar .
 \end{gather}
The imaginary part of \eqref{www} is a phase that suppress the amplitude everywhere but where the argument is zero or a multiple of $2\pi$. This gives the condition
\begin{gather*}
\textrm{Re}\, \tilde z = -\fracs32
 \lambda \vol_o j^{\frac12}  ,
\end{gather*}
that, together with the  condition \eqref{Re}, becomes
\begin{gather*} 
\textrm{Re}\,\tilde z =- \fracs32\lambda \vol_o j_o^{\frac12} = - \fracs32\lambda \vol_o \sqrt{{ \textrm{Im}\,\tilde z}/{4t\hbar}} .
\end{gather*}
This expression yields the Friedmann equation:
recall that $\textrm{Re}\,\tilde z \sim \dot a$ and $\textrm{Im}\,\tilde z \sim a^2$ so that,
squaring the previous equation, we obtain
\begin{gather*}
  \left(\frac{\dot{a}}{a}\right)^2=\frac{\Lambda}{3},
\end{gather*}
where $\Lambda= 27 \lambda^2\vol_o^2/16 t\hbar$.
The same result can be obtained by a dif\/ferent technique: the transition amplitude results to be annichilated by a Hamiltonian constraint. In the classical limit, this is
\begin{gather*}
    \big( \tilde z+ \fracs32  \lambda \vol_o j_o^{\frac12}\big)^2+\overline{\big( \tilde z+ \fracs32  \lambda \vol_o j_o^{\frac12}\big)^2}=0
\end{gather*}
that gives
\begin{gather*}
i4\, \textrm{Im}\, \tilde z  \big(\textrm{Re}\, \tilde z + \fracs32 \lambda \vol_o j_o^{\frac12}\big)=0 .
\end{gather*}

Notice that
we don't obtain the curvature term $k/a^2$ in the full Friedmann equation
\begin{gather}
  \left(\frac{\dot{a}}{a}\right)^2=\frac{\Lambda}{3} - \frac k{a^2} .
        \label{ds}
\end{gather}
This is because of the approximation taken in the evaluation of the gaussian sum. Since we ask for large~$j$, namely for a large distance regime, the curvature term is neglected being a higher order in~$j$. Finding a way to relax this approximation is an urgent issue in spinfoam cosmology: the higher order in $j$  would in fact provide us also the f\/irst quantum corrections. Christian
R\"oken has observed that equation~\eqref{ds}, including the curvature term, can be obtained simply by keeping the~$+1$ term of the $(j_\ell +1)$ in~\eqref{www} and then rescaling $a$ appropriately~\cite{Roken:2012}.

\subsection{Graph independence and 2-node model}

As mentioned above, spinfoam cosmology was introduced in \cite{Bianchi:2010zs} with a calculation based on the dipole graph with 4 links. This choice was dictated by simplicity and by the fact that this graph, already studied in the canonical context, has a compelling interpretation being a triangulation of the 3-sphere. But the computation above shows that, studying the large distance limit the results for the dipole are the same as for other regular graphs. (We emphasize the fact that numerical investigation shows that the large $j$ convergence is very fast, and the asymptotic regime is already essentially reached with $j\sim 3$.)
Let us discuss the terms in  \eqref{www} that carry a~dependence on the graph used.

At the stationary point the Hessian $\mathrm{H}$ give a contribution $N_{\Gamma}$ that depends
 on the graph $\Gamma$ trough its numbers of links $L$ and nodes $N$,  and a characteristic term $j_o^{-3}$ that is independent of the graph. This is the norm squared of the Livine--Speziale coherent regular cell of size~$j_o$~\cite{Livine:2007hc} (recently calculated for the Lorentzian signature \cite{Bianchi:2011ta}). Notice that since we have f\/ixed the normals, degenerate contributions are not allowed (being these present, we would have had further terms $\sim j_o^{-1}$).

The volume $\vol_o$ depends on the graph used. On the other hand, such a  cosmological-constant term has been introduced as an edge amplitude. This edge amplitude can be viewed as a~redef\/inition of the vertex. Possible normalization ambiguities, coming from the introduction of this term, can therefore be absorbed in the vertex amplitude~\cite{Magliaro:2010vn}.

The transition amplitudes that we are dicussing are in fact  not normalized. The arbitrary normalization of the vertex amplitude is f\/ixed by cylindrical consistency \cite{Magliaro:2010vn}.
We f\/ind that the dependence on the number of nodes enters only in the term~$N_\Gamma$ in~\eqref{eccola}, and it can be counterbalanced by normalizing appropriately the amplitude.
This implies that this result can be obtained also in the 2-node model, with the only caveat that $N_\Gamma$ would be the one for 2 nodes.

Let us consider now the number of links.
In the semiclassical limit the expression of the amplitude can be given in the form
\begin{gather*}
W(\tilde z)= \left(2j_o \sqrt{{\f\pi t}} e^{-\f{\tilde z^2}{8t\hbar}}  \right)^{L}  \f{N_\Gamma}{j_o^{3}}  .
\end{gather*}
Here the information about the semiclassical dynamics is coded in the kernel of the exponential, that does not depends on the number of links.

Using this and \eqref{Re}, we conclude
\begin{gather*}
W(\tilde z)= N \tilde z^{L-3} e^{-\f{L}{2t\hbar}\tilde z^2},
\end{gather*}
where $N=(\fracs{4\pi}{t})^{L/2} (\f{-i}{4t\hbar} )^{L-3}  N_{\Gamma}$. Finally, inserting into~\eqref{factoriz} we have
\begin{gather}
W(\tilde z_\ini,\tilde z_\fin)= N^2   (\tilde z_\ini \tilde  z_\fin)^{L-3}  e^{-\f{L}{2t\hbar}(\tilde z_\ini^2+\tilde z^2_\fin)} .
\label{eccola}
\end{gather}

This is the transition amplitude between two cosmological homogeneous isotropic coherent states, with $N$ and $L$ links such that the graph is \emph{regular} (i.e.\ every node has the same valency). Notice that there are an inf\/inite number of such graphs. For instance, two nodes can be connected by arbitrary number of links.  Examples of regular graphs with $N>2$ are given by the (dual of) the Platonic solids.

The result of this calculation is that the support of the transition amplitude, obtained trough the conditions on the real and the imaginary part of~$\tilde z$ that yields the Friedmann equation, is not sensitive to the number of links or the number of nodes of the graph used.

One aspect of the problem that (to the best of our knowledge) has not yet been studied is to compare the transition amplitude for homogeneous to homogeneous geometries to the transition amplitude from homogeneous to inhomogeneous geometries on the same graph. Namely to understand wether the quantum evolution smears out a state that is peaked on a homogeneous geometry to one roughly equally distributed over all geometries.

\subsubsection[Covariant $\U(N)$ framework]{Covariant $\boldsymbol{\U(N)}$ framework}

In Section~\ref{section3}, we have illustrated the power of the $\U(N)$ framework,
for instance for the implementation of homogeneity and isotropy in the
2-node model. Is it possible to use this technique in the covariant
theory illustrated in this section?  Work in this directions is still
under development. A f\/irst step has been recently proposed in~\cite{Livine:2011up}, where $\U(N)$ coherent states are def\/ined on the
dipole (Fig.~\ref{2vertex}) and a simple $\SU(2)$ transition amplitude between
these states is studied. This is an interesting research direction
that deserves further studies. In particular, the next step should be
to include  in the picture the full Lorentzian spinfoam dynamics of
general relativity.

\section{Summary}\label{section5}

In the early days of LQG, the realization that the theory lead to Planck scale discreteness nourished the intuition that the only way of recovering a continuous space from the theory was to have a very large number of links~\cite{Ashtekar:1992tm,Iwasaki:1992qy}.  Gaining clarity about the distinction between large number of links and large quantum numbers such as the area quantum number~$j$, lead to the realization that the theory can describe large semiclassical geometries also over a small number of nodes and links. Here we have reviewed a number of constructions in Loop Quantum Gravity, based on the idea of truncating the Hilbert space of the theory down to the states supported on a simple graph with two nodes~\cite{Rovelli:2008ys}.

The restriction of the full LQG Hilbert space to a simple graph is a
truncation of the degrees of freedom of the
full theory. It def\/ines an approximation where concrete calculations
can be performed. The approximation
is viable in  physical situations where only a small number of the
degrees of freedom of General Relativity
are relevant. A characteristic example is  cosmology.

The 2-node graphs (dipole) with 4 links def\/ines the simplest
triangulation of a 3-sphere and can accommodate
the anisotropic degrees of freedom of a Bianchi IX model, plus some
inhomogeneous degrees of freedom, which
can be seen as the lowest modes in a spherical-harmonic expansion,
following a technique introduced by Regge and Hu.
In this context, a Bohr--Oppenheimer
approximation provides a tool to separate heavy and light degrees of
freedom, and extract the FLRW dynamics.
This way of deriving quantum cosmology from LQG is dif\/ferent from the
usual one: in standard loop
quantum cosmology, the strategy is  to start from a symmetry-reduced
system, and quantize the single or the
few degrees of freedom that survive in the symmetry reduction. Here
instead we consider a truncated version of
the full quantum theory of gravity, in the LQG framework, and look for
a ``cosmological sector'' inside the theory.

In Section~\ref{section3} we have stepped up to a 2-node graphs with arbitrary
number of links. This system provides an
immediate application of the $\U(N)$ formalism~\cite{Borja:2010gn}.
 This formalism is based on the observation that
the LQG Hilbert space of intertwiners with $N$-legs and f\/ixed area is
an irreducible representation of
the group $\U(N)$ \cite{Freidel:2009ck}.  The relation is made
explicit using  the  Schwinger
representation of $\SU(2)$. Furthermore, the full state space of
$N$-leg intertwiners can be endowed
with a Fock-space structure, with annihilation and creation operators~$F_{ij}$ and~$F_{ij}^\dagger$.

In the 2-node graph context, one can def\/ine operators $e_{ij}$ and
$f_{ij}$, that are $\SU(2)$ invariant and consistent with the
matching conditions between the intertwiners (ensuring that the
spin number of one leg is the same seen from both nodes).
The system has a $\U(N)$ global symmetry, given by a generalization of
the matching conditions.
The space of the states invariant under this symmetry is homogeneous and
isotropic. This construction def\/ines the homogeneous and isotropic
conf\/igurations
via a symmetry reduction at the quantum level, and may shed  further
light on the
relation between Loop Quantum Cosmology and Loop Quantum Gravity.
In particular, the construction leads to the def\/initions of a
nontrivial consistent Hamiltonian operator
for the homogeneous/isotropic sector, which has intriguing
mathematical analogies with the operators
used in LQC.

We have also reviewed the classical spinor system whose quantization
yields the Hilbert space of intertwiners of LQG
\cite{Borja:2010rc,Livine:2011gp}.  This framework permits the
construct of the classical counterpart of the $\U(N)$
Hamiltonian for the 2-node model, def\/ining an ef\/fective classical
dynamics for this system. The equations of motion of this classical
system can be solved, and the resulting dynamics shows analogies
with the results of LQC.

In Section~\ref{section4} we switched to the covariant, or spinfoam, def\/inition of
the dynamics. Here the dynamics isnot def\/ined by a Hamiltonian, but
rather directly by a transition amplitude between two states of the
quantum geometry. At the f\/irst order in the vertex expansion, this
amplitude factorizes and def\/ines a ``wave function of the universe''
\`a la Hartle--Hawking.  In the classical limit, the amplitude turns
out to be peaked on the solutions of Einstein equations, that is, In
presence of isotropy and homogeneity, of the Friedmann equation~\cite{Bianchi:2010zs}. The model can include the presence of the
cosmological constant~\cite{Bianchi:2011ym}. This is obtained by
inserting a ``face amplitude'' term into the spinfoam amplitude and
can be seen as an \emph{effective} way to include~$\Lambda$. It is
generally thought that at the fundamental level the cosmological
constant should emerge in a quantum deformed version of the spinfoam
theory. Such a quantum deformation should hopefully results in a term
that match with the one that we have heuristically introduced.

The amplitude for  homogeneous isotropic states was f\/irst computed
using the dipole graph, but, remarkably, the classical limit of the
amplitude turns out to be independ on the (regular) graph chosen~\cite{Vidotto:2011qa}. This result supports the viability of the
approximation taken by restricting the theory to a single graph.
States of large regular graphs include in principle inhomogeneous
quantum f\/luctuations, beyond perturbations techniques as usually
utilized in quantum cosmology.

In closing, we point out three directions where the techniques
reviewed here might turn out to be
useful to better understand loop cosmology.

First, the $\U(N)$ symmetry provides an elegant way to impose
inhomogeneity and anisotropy and
so far the attention has focused on the $\U(N)$-invariant states. Can
we go beyond this sector?
Indeed, the action that def\/ines the full classical kinematics and dynamics of
spinnetwork states on the 2-node graph is a nontrivial matrix
model def\/ined in terms of the unitary matrices~$\Ua$ and~$\Ub$,
with quartic interaction terms. It would be very interesting to see
what kind of anisotropy does our model describe in the context of
loop cosmology.

Second, the relation between the dif\/ferent dynamics def\/ined by
standard LQC, by the
Hamiltonian of the $\U(N)$ framework and by the spinfoam amplitude need to be
compared in detail. For this, in particular, he analysis of the
spinfoam amplitude should
be developed beyond the semiclassical limit.

Finally, all the analysis reviewed here is in the context of pure
gravity, and disregard the
presence of matter.  The coupling of fermions and Yang--Mills f\/ields is
simple and natural in
the Hilbert space of Loop Quantum Gravity at the kinematical level. In
the $\U(N)$ approach
there is a direct formulation of this coupling in  spinor phase space
before quantization.

The coupling of fermions to canonical LQG is well understood \cite{Baez:1997bw,MoralesTecotl:1994ns,MoralesTecotl:1995jh,Thiemann:2007zz,Thiemann:2002nj}.
In the spinfoam approach, the dynamical coupling with fermions and
Yang--Mills f\/ields has
been def\/ined recently in \cite{Bianchi:2010fk,Han:2011as}
and has not yet been much studied.
Including matter couplings
is clearly essential for understanding the quantum dynamics of cosmology.

\subsection*{Acknowledgements}

This work was in part supported by the Spanish MICINN research
grants FIS2008-01980 and FIS2009-11893. IG is supported by the
Department of Education of the Basque Government under the
``Formaci\'{o}n de Investigadores'' program.

\pdfbookmark[1]{References}{ref}
 \LastPageEnding

\end{document}